\pdfoutput=1
\documentclass{article}
\PassOptionsToPackage{numbers}{natbib}
\usepackage[final]{neurips_2024}

\usepackage[utf8]{inputenc} 
\usepackage[T1]{fontenc}    
\usepackage{hyperref}       
\usepackage{url}            
\usepackage{booktabs}       
\usepackage{amsfonts}       
\usepackage{nicefrac}       
\usepackage{microtype}      
\usepackage{xcolor}         
\usepackage{overpic}
\usepackage{graphicx}
\usepackage{amsmath}
\usepackage{algorithm}
\usepackage{algpseudocode}
\newcommand*{\boldone}{\text{\usefont{U}{bbold}{m}{n}1}}

\title{Model Decides How to Tokenize: Adaptive DNA Sequence Tokenization with MxDNA}

\author{%
  Lifeng Qiao$^{1,2}$\thanks{Work done during an internship at Shanghai Artificial Intelligence Laboratory.}   , Peng Ye$^{1,3}$\thanks{Corresponding Author.}   , Yuchen Ren$^{1,4}$,  Weiqiang Bai$^{1}$, Chaoqi Liang$^{1}$, \\
  \textbf{Xinzhu Ma}$^{1,3}$, \textbf{Nanqing Dong}$^{1}$, \textbf{Wanli Ouyang}$^{1,3}$ \\
  $^1$Shanghai Artificial Intelligence Laboratory, $^2$Shanghai Jiao Tong University \\
  $^3$The Chinese University of Hong Kong, $^4$The University of Sydney \\
  \texttt{yepeng@pjlab.org.cn}
}

\begin{document}
\maketitle

\begin{abstract}
Foundation models have made significant strides in understanding the genomic language of DNA sequences. However, previous models typically adopt the tokenization methods designed for natural language, which are unsuitable for DNA sequences due to their unique characteristics. In addition, the optimal approach to tokenize DNA remains largely under-explored, and may not be intuitively understood by humans even if discovered. To address these challenges, we introduce MxDNA, a novel framework where the model autonomously learns an effective DNA tokenization strategy through gradient decent. MxDNA employs a sparse Mixture of Convolution Experts coupled with a deformable convolution to model the tokenization process, with the discontinuous, overlapping, and ambiguous nature of meaningful genomic segments explicitly considered. On Nucleotide Transformer Benchmarks and Genomic Benchmarks, MxDNA demonstrates superior performance to existing methods with less pretraining data and time, highlighting its effectiveness. Finally, we show that MxDNA learns unique tokenization strategy distinct to those of previous methods and captures genomic functionalities at a token level during self-supervised pretraining. Our MxDNA aims to provide a new perspective on DNA tokenization, potentially offering broad applications in various domains and yielding profound insights. Code is available at \url{https://github.com/qiaoqiaoLF/MxDNA}.

\end{abstract}

\section{Introduction}
\label{introduction}
Foundation models in natural language processing (NLP) have achieved remarkable success, transforming how machines understand and generate human language~\cite{devlin2018bert,brown2020language,touvron2023llama}. Inspired by this success, researchers are now exploring the application of foundation models to decode the complex ``language'' of genomic sequences, aiming to potentially revolutionize our understanding of genomics~\cite{ji2021dnabert,zhou2023dnabert,dalla2023nucleotide,nguyen2024hyenadna}. Tokenization, a critical initial step in NLP models, leverages human knowledge of natural language structures such as grammar and punctuation to segment text into meaningful units. However, DNA sequences present a distinct challenge: they lack natural delimiters and their ``grammar'' is not readily understood by humans. These challenges make the tokenization process of DNA sequences not straightforward.

\begin{figure}[ht]
  \centering
  \includegraphics[width=\textwidth]{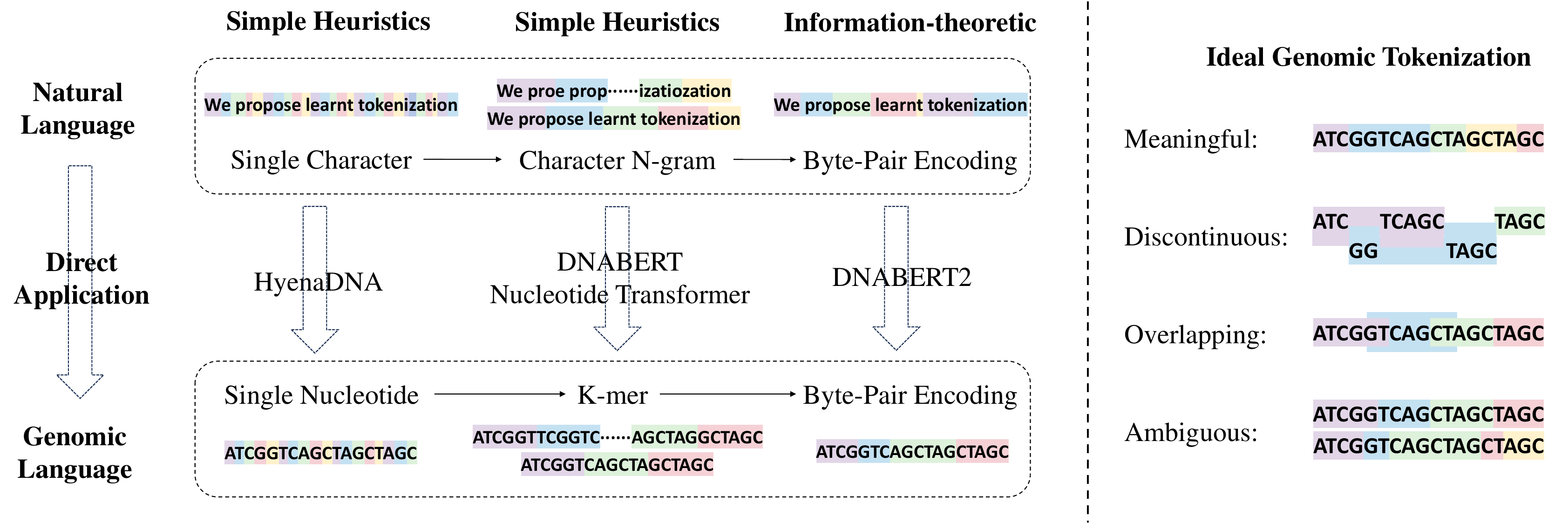}
  \caption{Evolution of tokenization and Ideal Properties. \textbf{Left:} The progression from basic tokenization methods to more sophisticated techniques, with the direct but unsuitable applications from natural language to genomic language. \textbf{Right:} the ideal tokenization properties for genomics—Meaningful, Discontinuous, Overlapping, and Ambiguous—outlined in~\cite{vu2023linguistically}, which our MxDNA aims to achieve.}
  \label{fig:motivations}
\end{figure}

Various tokenization methods have been employed by existing foundation models to analyse DNA sequences~\cite{nguyen2024hyenadna,ji2021dnabert,dalla2023nucleotide,zhou2023dnabert}. For example, single nucleotide tokenization~\cite{nguyen2024hyenadna} treats each nucleotide as an individual token, K-mer~\cite{ji2021dnabert,dalla2023nucleotide} segments the DNA blocks of k consecutive nucleotides, and Byte-Pair Encoding (BPE)~\cite{zhou2023dnabert} iteratively merges the most frequent pairs of existing tokens. All of these methods are borrowed directly from NLP as depicted in Fig.~\ref{fig:motivations}, each with its own inherent limitations. Single nucleotide tokenization, while offering high resolution for input, leads to an extremely large number of tokens, significantly increasing the complexity of the model. K-mer comes in two forms: overlapping and non-overlapping. Overlapping K-mer, despite its attempt to capture more contextual information, does not offer substantial benefits over single nucleotide approaches and can suffer from information leakage~\cite{zhou2023dnabert, liang2023rethinking}. Non-overlapping K-mer greatly reduces tokenized sequence length but can disrupt a potentially meaningful unit by splitting it into separate K-mers. BPE, adopted from NLP, attempts to optimize vocabulary size but often results in suboptimal segmentation that may not correspond to meaningful units~\cite{bostrom2020byte, hofmann2021superbizarre}.

Unlike natural languages, where linguistically meaningful units such as words and sentences are almost standardized and well understood, the optimal approach to tokenize DNA remains under-explored due to the complex and varied nature of genomics. In NLP, common tokenization strategies have been validated by human knowledge, but such understanding does not extend to DNA. Consequently, rather than relying manually crafted tokenization rules, it may be better to trust a neural network to learn and determine the most effective tokenization strategy for genomic sequences. Additionally, recent research suggests that biologically meaningful protein tokens can be discontinuous, overlapping, and may require mapping to several tokenization possibilities~\cite{vu2023linguistically,robert2022unconstrained,akbar2021compact}, properties that are likely applicable to DNA sequences due to the genetic central dogma~\cite{crick1970central}. To handle these complexities,
we can further equip our model with capabilities to manage discontinuities, overlaps, and the ambiguities of genomic sequences explicitly. 

Building on the analysis above, we introduce MxDNA (``Mx'' draws from \textbf{M}ixture of \textbf{Ex}perts~\cite{shazeer2017outrageously}),
a novel framework designed to autonomously learn an effective DNA tokenization strategy solely through gradient decent. The core of the framework starts with a sparse Mixture of Convolution Experts that identifies and embeds basic units within DNA sequences. Unlike conventional Mixture of Experts models, which focus on scaling up the model while maintaining computational efficiency~\cite{shazeer2017outrageously,fedus2022switch,jiang2024mixtral}, the experts in MxDNA are uniquely designed to capture DNA basic units of varied lengths. Following this, a deformable convolution~\cite{dai2017deformable,zhu2019deformable} assembles these basic units into final tokens. Throughout the process, MxDNA is explicitly equipped to manage the inherent discontinuities, overlaps and ambiguities in genomic sequences, enabling it to handle complex biological characteristics it encounters. Furthermore, we incorporate a cross-attention mechanism to align the output resolution with the original input during pretraining on a masked language modeling task~\cite{devlin2018bert}.

The proposed MxDNA demonstrates strong performance on both the Nucleotide Transformer Benchmarks~\cite{dalla2023nucleotide} and the Genomic Benchmarks~\cite{grevsova2023genomic}. Despite only being pretrained on human reference genome, it still outperforms or matches previous models~\cite{nguyen2024hyenadna,ji2021dnabert,dalla2023nucleotide,zhou2023dnabert}—some pretrained on multi-species data and for much longer duration, achieving state-of-the-art average performance and the best on 15 of the 26 individual tasks. Finally, by visualizing the learnt tokenization process, we illustrate that MxDNA learns unique tokenization strategy distinct to those of previous methods and captures genomic functionalities at a token level during self-supervised pretraining, potentially offering novel biological insights. Our contributions can be summarized as follows:

\begin{itemize}
 \item \textbf{Learnt Tokenization}: We highlight the unsuitability of current DNA tokenization methods directly borrowed from NLP. Based on the belief that humans may not know the best tokenization approach but a model could potentially discover it, we propose a novel approach where the model autonomously learns an effective tokenization strategy.

  \item \textbf{Architectural Design}: We introduce a sparse Mixture of Convolution Experts coupled with a deformable convolution to dynamically learn tokenization, specifically designed to manage the inherent discontinuities overlaps and ambiguities in genomic sequences. Additionally, we leverage cross attention to align input and output sequence length to enable self-supervised pretraining. 
  
  \item \textbf{Empirical Results}: MxDNA demonstrates robust quantitative performance  with less pretraining data compared to some existing models, achieving state-of-the-art average performance on both Nucleotide Transformer Benchmarks and Genomic Benchmarks. Furthermore, visual analysis of the tokenization behaviour and the token embedding space highlights the unique strategy and capability to capture genomic functionalities at a token level of MxDNA, potentially offering new biological insights.
  
\end{itemize}
\section{Background and Related Work}
\label{related_works}

\subsection{Tokenization Methods} 

Tokenization is a fundamental step in both natural language processing (NLP) and DNA sequence modelling, transforming complex texts or DNA sequences into manageable tokens. In NLP, whitespace tokenization uses spaces and punctuation as delimiters but faces out-of-vocabulary issues. Similarly, in both fields, character (or single nucleotide in DNA) tokenization provides high resolution but can lead to computational inefficiency~\cite{kim2016character, al2019character, nguyen2024hyenadna, schiff2024caduceus}. N-gram in NLP and K-mer in DNA analysis both use contiguous sequences of N (K) items from given inputs~\cite{brown1992class, mcnamee2004character, dalla2023nucleotide, zhang2023dnagpt, ji2021dnabert}, but can disrupt meaningful units due to their fixed-length nature (non-overlapping) or lead to potential information redundancy or leakage (overlapping)~\cite{zhou2023dnabert,liang2023rethinking}. Byte-Pair Encoding (BPE) is employed across both domains to reduce vocabulary size by merging frequent pairs of existing tokens~\cite{sennrich2015neural, zhou2023dnabert}. However, it might not adequately encode more complex patterns and are unreliable for finding linguistically sound tokens~\cite{hofmann2021superbizarre,bostrom2020byte}. These rule-based methods show limitations in different aspects, and our study aim to develop a learning-based tokenization method without these limitations.

\subsection{DNA Foundation Models}

Recent advancements in DNA modeling have leveraged foundation models to decode the complex language of genomes. \textbf{DNABERT}~\cite{ji2021dnabert} pioneers the use of a BERT-like pretrained model for genomic sequence analysis, enhancing the understanding of nucleotide relationships via attention mechanisms. \textbf{Nucleotide Transformer}\cite{dalla2023nucleotide} offers a comprehensive analysis of foundation models pretrained on DNA sequences, with model sizes reaching up to 2.5 billion parameters and pretraining data drawn from the 1000G\cite{byrska2022high} human genomes and 850 various species. \textbf{DNABERT2}~\cite{zhou2023dnabert} introduces an enhanced genome foundation model, utilizing an efficient BPE tokenizer and techniques to address input length constraints, resulting in reduced time and memory consumption while improving performance. \textbf{HyenaDNA}~\cite{nguyen2024hyenadna} introduces a genomic foundation model capable of handling context with 1 million tokens at single nucleotide resolution, enabling the first exploration of in-context learning in genomics. \textbf{DNAGPT}~\cite{zhang2023dnagpt} extends the traditional GPT model by integrating tasks such as binary classification of DNA sequence order and numerical regression for predicting guanine-cytosine content, alongside developing a comprehensive token language. \textbf{Caduceus}~\cite{schiff2024caduceus} designs an architecture that leverages the long-range Mamba~\cite{gu2023mamba} block to support bi-directionality and reverse complementarity equivariance, addressing specific challenges in genomic analysis. Following VQVAE~\cite{van2017neural}, \textbf{VQDNA}~\cite{li2024vqdna} employs a convolutional encoder alongside a vector-quantized codebook to model tokenization, sharing a similar motivation with us yet ultimately adopting distinct solutions. Each work offers unique insights and innovations to the filed. Our research specifically concentrate on the tokenization methods for DNA, hopefully providing our unique contributions to the filed.

\section{Method}
\label{method}

\begin{figure}[ht]
  \centering
  \includegraphics[width=\textwidth]{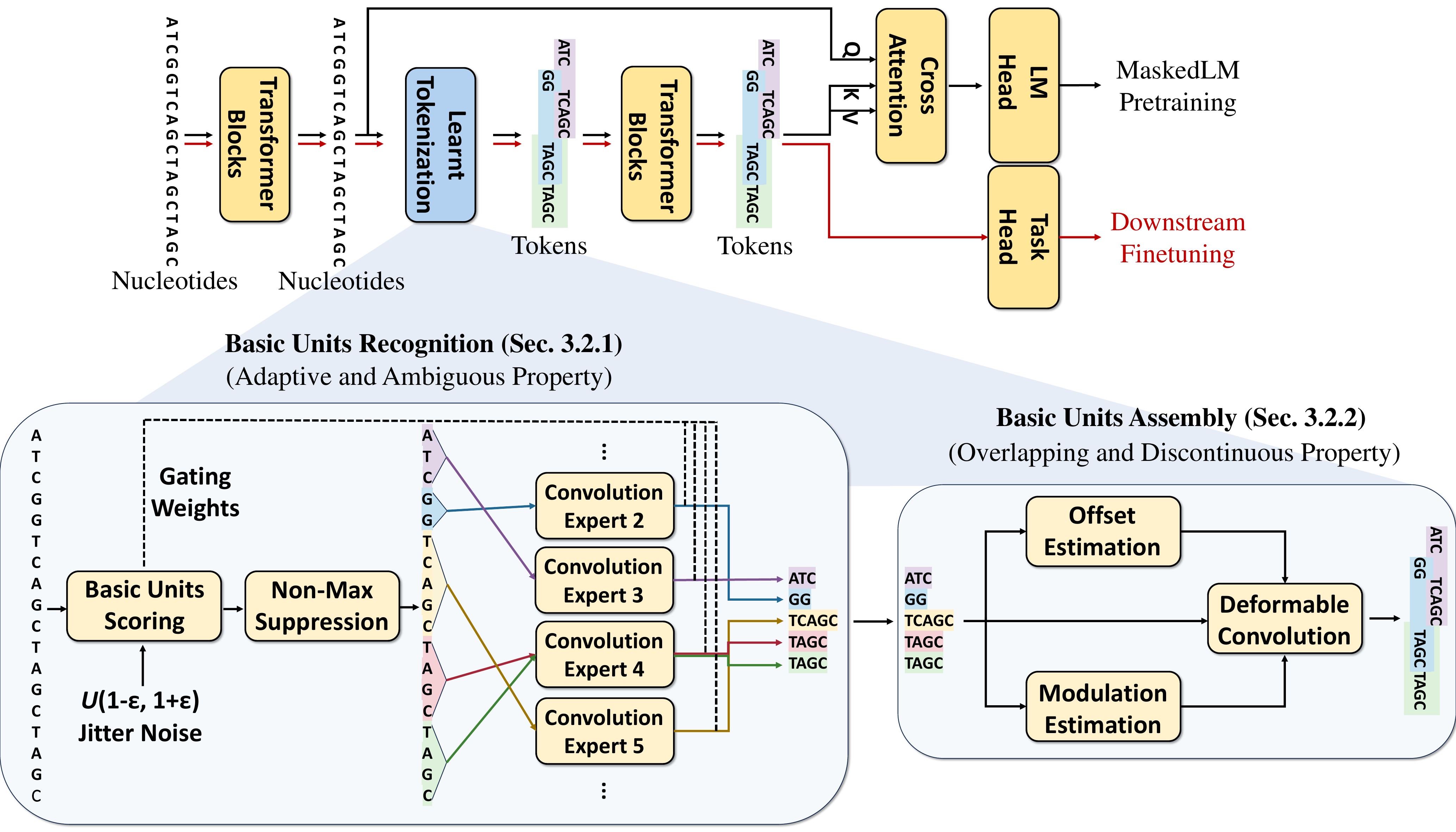}
  \caption{
Our proposed MxDNA. \textbf{(Top) Overall pipeline of the MxDNA model:} Black arrows indicate pretraining data flow, and red arrows indicate finetuning data flow. The learnt tokenization module tokenizes single nucleotide input into learnt tokens.
\textbf{(Bottom) Illustration of the learnt tokenization module:} Meaningful basic units are recognized with a linearly scoring layer and non-maximum suppression, embedded through convolution experts (Sec.~\ref{Recognition}), and assembled into final tokens by a deformable convolution. (Sec.~\ref{Assembly}) This process ensures meaningful, discontinuous, overlapping, and ambiguous tokenization, addressing the unique properties of genomic data.}
\end{figure}

\subsection{Motivation}

The concept of ``correct'' tokenization in genomic sequences analysis remains undefined due to the complex nature of genomics. Unlike natural languages, where linguistically meaningful units are well-understood, biological units in genomics are not limited to contiguous nucleotide or amino acid sequences. Instead, they often encompass discontinuous, overlapping, and ambiguous segments crucial for understanding biological functions~\cite{vu2023linguistically}. Current DNA modeling practices directly borrow tokenization methods from natural language processing (NLP), such as single nucleotide tokenization, K-mer and Byte-Pair Encoding (BPE). These fixed, predefined approaches, though useful, often fail to capture the unique properties of DNA sequences, which lack explicit delimiters and consist of biologically meaningful units that defy simple segmentation.

Recognizing these challenges, MxDNA was developed based on the belief that although an optimal tokenization schema for genomic sequences is yet to be discovered, we can explicitly equip our model with the desired tokenization properties—such as handling discontinuities, overlaps, and ambiguities, and allow it to learn and adapt its tokenization strategy all by itself.

\subsection{Learnt Tokenization Module}

This section introduces our learned tokenization module, which is central to our approach. The module first identifies meaningful basic units within the input sequence, which are then assembled into tokens with discontinuous, overlapping, and ambiguous properties. Implementation details are in Appx.~\ref{AppendixA2}.

\subsubsection{Basic Units Recognition}
\label{Recognition}
\paragraph{Basic Units Scoring} Initially, MxDNA identifies the basic units that serve as the building blocks for tokens. It estimates the probability of the existence of various sized basic units centred at each nucleotide position by a linear gating mechanism commonly used in Mixture of Experts models. Following this, one-dimensional non-maximum suppression is applied to eliminate redundant proposals and select the most significant basic units.

Specifically, given the input nucleotide sequence \(X \in \mathbb{R}^{l \times d}\), where \(l\) is the sequence length and \(d\) is the hidden dimension,  \(X\) is first linearly scored to produce \(S \in \mathbb{R}^{l \times n}\), where \(n\) represents the number of experts. Training-time multiplicative jitter noise~\cite{fedus2022switch} is applied to introduce \textbf{ambiguity}, while ensuring deterministic inference. The jitter noise is applied by multiplying the scores with a random factor sampled uniformly between \([1-0.01, 1+0.01]\), resulting in slight perturbations to the probability distribution used for tokenization. 

Modified non-maximum suppression is then applied  to \(S\), where \(S_{ij}\) indicates the presence probability of a basic unit of length \(L_j\) centered at position \(i\), and \(L \in \mathbb{N} ^ n\) is a predefined set of lengths. The results are tracked using an expert mask \(M \in \mathbb{N}^l\), where each \(M_i\) is a natural integer indicating the presence of a basic unit's center of length \(M_i\) at position \(i\).

\paragraph{Basic Units Embedding} After identifying the basic units, the nucleotides within each unit are aggregated to form embeddings. Convolution kernels of corresponding sizes are applied to the center of each basic unit to capture local features. The initial scoring and gating into specific convolution experts is similar to the Mixture of Experts paradigm,  with each expert being a convolutional unit focusing on a specific segment rather than a single nucleotide.

Specifically, a basic unit at position \(i\) of length \(L_j = M_i\) is processed by the convolution expert \(E_j\) with kernel size \(L_j\), and weighted by \(\text{softmax}(S_i)_j\), aggregating the nucleotides within the unit. This transforms the original input \(X \in \mathbb{R}^{l \times d}\) into an array of basic units \(U \in \mathbb{R}^{l \times d}\), where \(k\) is the number of basic units:

\begin{equation}
\label{SMoCEeq}
U_i =
\begin{cases}
E_j(X_{\left[i-\left \lceil\frac{M_i}{2}\right \rceil+1:i+\left \lfloor\frac{M_i}{2}\right \rfloor\right]}) \cdot \text{softmax}\left(S_i\right)_{j}, \text{where } L_j = M_i &, M_i > 0 \\
0 &, M_i = 0
\end{cases}
\end{equation}

Then, the unwanted entries \(\left\{ i | M_i = 0 \right\}\) of \(U\) are removed to keep the basic units \(U \in \mathbb{R}^{k \times d}\) only, where \(k\) is the number of basic units.

\subsubsection{Basic Units Assembly}
\label{Assembly}

\paragraph{Distal Relation Estimation} Building upon the identified basic units, the more complex genomic patterns that extend beyond simple segmentation are modelled by a one-dimensional deformable convolution. This technique uniquely accommodates the modeling of complex local geometric transformations, adaptively adjusting to the input sequence. The linkages between the distal basic units are modeled by the offsets and modulation factors of each basic unit.

Following~\cite{dai2017deformable, zhu2019deformable}, offsets \(\Delta P \in \mathbb{R}^{k \times f}\) and modulation factors \(\Delta M \in \mathbb{R}^{k \times f}\) are computed based on the basic units \(U\) to model the distal relationships among them. This strategy ensures the combination of basic units is \textbf{discontinuous}, and reuses units across tokens achieve the \textbf{overlapping} property.

\paragraph{Final Tokens Embedding}  Using the computed offsets and modulation factors, deformable convolution is applied to embed basic units into final tokens. The embedding process for each position incorporates deformations of the convolution kernel specified by the offsets, with the results modulated by the modulation factors.

Specifically, a one-dimensional deformable convolution with kernel size \(f\) is applied to embed these basic units into the final learnt tokens \(T \in \mathbb{R}^{k \times d}\):

\begin{equation}
\label{DCNeq}
T_i = \sum_{p \in \left\{-\left \lceil\frac{f}{2}\right \rceil + 1,...,\left \lfloor\frac{f}{2}\right \rfloor\right\}} w_p \cdot U_{i+p+\Delta p} \cdot \Delta m
\end{equation}
For a fractional location \(p' = i+p+\Delta p\), bilinear interpolation is implemented as:
\begin{equation}
U_{p'} = \sum_{q \in \left\{1,...,k\right\}} \max\left(0, 1 - \left|p' - q\right|\right) \cdot U_q
\end{equation}

\subsection{Overall Pipeline}

The framework begins with single nucleotide input represented as \(X_{input} \in \mathbb{R}^{l \times d}\). This single nucleotide resolution input allows for fine-grained analysis of genomics from the beginning.

Initially, \(X_{input}\) is processed through several transformer encoder blocks designed to extract global relationships within the sequence, producing \(X \in \mathbb{R}^{l \times d}\). This sets the stage for effective tokenization. Following this, the learnt tokenization module transforms the nucleotide sequence \(X\) into a more manageable form \(T \in \mathbb{R}^{k \times d}\), improving the efficiency and focus of subsequent layers. The tokenized output \(T\) is then passed through another series of transformer encoder blocks to further refine the token representation to \(T_{output} \in \mathbb{R}^{k \times d}\), enhancing the model's ability to encode deeper genomic information.

For the mask language modeling pretraining stage, the enriched nucleotide level representation \(X\) serves as the query, with the refined tokenized output \(T_{output}\) acting as both the key and value. This setup maps the output resolution to single nucleotides, essential for reconstructing masked tokens. During the finetuning stage, the [CLS] token of \(T_{output}\) is used for classification by convention.

\section{Experiments}
\label{experiments}
In this section, we first introduce the implementation and pretraining settings of MxDNA. Then, we evaluate MxDNA against other foundation models on Genomic Benchmarks~\cite{grevsova2023genomic} and Nucleotide Transformer Benchmarks~\cite{dalla2023nucleotide}. Next, we present ablation studies on the effect of different tokenization methods and different components of MxDNA. Finally, we conduct a simple analysis on the tokenization behaviors of MxDNA. Experiment settings and results are detailed in Appx.~\ref{AppendixExp}.

\subsection{Model Implementation \& Pretraining}

Our MxDNA is built on the architecture Nucleotide Transformer v2 100M model with 512 hidden units and 22 layers, totaling approximately 100M parameters. Specifically, the model's learnt tokenization module includes 10 convolution experts with kernel sizes ranging from 1 to 10, along with a deformable convolution block with a kernel size of three. We integrate this module by replacing the fifth transformer block, aiming to avoid introducing additional computations.

MxDNA is pretrained on the whole Human Reference Genome~\cite{HG38} on masked language modeling task~\cite{devlin2018bert} with 15\% of the nucleotides randomly masked. An auxiliary balancing loss with a weight of 0.01 is used to prevent degradation towards a single expert, following~\cite{fedus2022switch}. The model undergoes training for 500k steps for main performance comparisons and 100k steps for ablations.

\subsection{Downstream Evaluation}

We primarily follow the evaluation settings of HyenaDNA~\cite{nguyen2024hyenadna}, performing evaluation on Nucleotide Transformer Benchmarks and Genomic Benchmarks. To ensure fair comparison, we fully finetune all the BERT-like DNA foundation models including Nucleotide Transformer v2 \cite{dalla2023nucleotide}, DNABERT~\cite{ji2021dnabert}, DNABERT2~\cite{zhou2023dnabert}, MxDNA under same hyperparameter settings. For HyenaDNA, we utilize the hyperparameters recommended by \cite{nguyen2024hyenadna, schiff2024caduceus}. All experiments are repeated with three random seeds, and we report the average performance with sample standard deviations.
 ~\footnote{The Nucleotide Transformer and DNABERT2 are pretrained on much larger datasets than other models.}

\begin{table}
  \caption{Genomic Benchmarks. Average performance across three random seeds for Nucleotide Transformer v2 100M, DNABERT, DNABERT2, HyenaDNA and MxDNA with sample standard deviations. We highlight the best values in \textbf{bold} type and \underline{underline} the second best.}
  \label{performace GB}
  \centering
  \scalebox{0.93}{
  \begin{tabular}{llllll}
    \toprule
    Dataset    & NTv2& DNABERT& DNABERT2 &HyenaDNA & MxDNA \\
    \midrule
    Average & 88.13 \tiny{\(\pm\) 0.03} & 87.50 \tiny{\(\pm\) 0.13} & \underline{88.29} \tiny{\(\pm\) 0.19} &87.17 \tiny{\(\pm\) 0.15}& \textbf{89.13} \tiny{\(\pm\) 0.13} \\
    \midrule
    Mouse Enhancers& \textbf{83.94} \tiny{\(\pm\) 0.41} &\underline{81.54} \tiny{\(\pm\) 0.86}  & 81.34 \tiny{\(\pm\) 0.84} &80.99 \tiny{\(\pm\) 0.72}&   80.57 \tiny{\(\pm\) 0.97}\\
    Coding vs Intergenomic&94.50 \tiny{\(\pm\) 0.06} &93.13 \tiny{\(\pm\) 0.05} &\underline{94.94} \tiny{\(\pm\) 0.34} &90.74 \tiny{\(\pm\) 0.11}&\textbf{95.28} \tiny{\(\pm\) 0.08}\\
    Human vs Worm&96.88 \tiny{\(\pm\) 0.18}&96.98 \tiny{\(\pm\) 0.07}& \underline{97.57} \tiny{\(\pm\) 0.04}  &96.53 \tiny{\(\pm\) 0.04}&\textbf{97.64} \tiny{\(\pm\) 0.01}\\
    Human Enhancer Cohn&74.33 \tiny{\(\pm\) 0.40}  &  74.54 \tiny{\(\pm\) 0.27}     &\textbf{75.93} \tiny{\(\pm\) 0.20} & 73.36 \tiny{\(\pm\) 0.15}& \underline{74.67} \tiny{\(\pm\) 0.09}\\
    Human Enhancer Ensembl&92.05 \tiny{\(\pm\) 0.38}& 92.18 \tiny{\(\pm\) 0.14}  & \underline{92.31} \tiny{\(\pm\) 0.03} &88.12 \tiny{\(\pm\) 0.17}& \textbf{93.13} \tiny{\(\pm\) 0.35}\\
    Human Regulatory&\underline{93.79} \tiny{\(\pm\) 0.12} & 88.16 \tiny{\(\pm\) 0.09} & 87.94 \tiny{\(\pm\) 0.54}&93.08 \tiny{\(\pm\) 0.22}& \textbf{94.11} \tiny{\(\pm\) 0.08}\\
    Human OCR Ensembl&78.51 \tiny{\(\pm\) 0.55} & \textbf{81.40} \tiny{\(\pm\) 0.11}  &80.94 \tiny{\(\pm\) 0.09} &79.15 \tiny{\(\pm\) 0.34}&\underline{81.05} \tiny{\(\pm\) 0.07}\\
    Human NonTATA Promoters&91.05 \tiny{\(\pm\) 0.47} &92.06 \tiny{\(\pm\) 0.20} & 95.34 \tiny{\(\pm\) 0.17}  & \underline{95.39} \tiny{\(\pm\) 0.26}&\textbf{96.56} \tiny{\(\pm\) 0.29}\\
    \bottomrule
  \end{tabular}}
\end{table}

\subsubsection{Genomic Benchmarks}

First, we begin our evaluation on the Genomic Benchmarks~\cite{grevsova2023genomic}, which consists of eight regulatory element classification tasks. For this benchmark, all BERT-like models are finetuned for 10 epochs with Top-1 accuracy reported for each dataset.

As shown in Table~\ref{performace GB}, MxDNA achieves the best performance on 5 out of 8 datasets and ranks in the top-2 on 7 out o
f 8 datasets. On average, MxDNA shows an improvement of 0.84 points compared to the second-best model, DNABERT2. These results demonstrate MxDNA's robustness and effectiveness in regulatory element classification.

\begin{table}
  \caption{Nucleotide Transformer Benchmarks. Average performance across three random seeds for Nucleotide Transformer v2 100M, DNABERT, DNABERT2, HyenaDNA and MxDNA with sample standard deviations. We highlight the best values in \textbf{bold} type and \underline{underline} the second best.}
    \label{performace NT}
  \centering
  \scalebox{0.93}{
  \begin{tabular}{llllll}
    \toprule
    Dataset    & NTv2 &DNABERT& DNABERT2 &HyenaDNA & MxDNA \\
    \midrule
    Average &70.70 \tiny{\(\pm\) 0.12} & 68.61 \tiny{\(\pm\) 0.16} & \underline{76.66} \tiny{\(\pm\) 0.20} & 75.96 \tiny{\(\pm\) 0.20}& \textbf{78.14} \tiny{\(\pm\) 0.12} \\
    \midrule
    Histone Markers Avg. & 55.22 \tiny{\(\pm\) 0.21} & 51.67 \tiny{\(\pm\) 0.17} & \underline{65.89} \tiny{\(\pm\) 0.46} & 65.24 \tiny{\(\pm\) 0.26}& \textbf{68.14} \tiny{\(\pm\) 0.19} \\
    \midrule
    H3& 78.22 \tiny{\(\pm\) 1.15} &77.41 \tiny{\(\pm\) 1.01}  & \underline{82.31} \tiny{\(\pm\) 0.22} & 80.86 \tiny{\(\pm\) 0.52}& \textbf{82.78} \tiny{\(\pm\)0.14}\\
    H3K14ac&51.76 \tiny{\(\pm\) 0.99}& 46.51 \tiny{\(\pm\)1.83} &65.13 \tiny{\(\pm\) 1.10} & \underline{65.96} \tiny{\(\pm\) 0.26}&\textbf{68.27} \tiny{\(\pm\)0.19}\\
    H3K36me3&59.18 \tiny{\(\pm\) 0.53} &50.98 \tiny{\(\pm\) 0.75}& \textbf{68.19} \tiny{\(\pm\) 0.82} &64.31 \tiny{\(\pm\) 0.33}& \underline{67.05} \tiny{\(\pm\) 1.05}\\
    H3K4me1&51.87 \tiny{\(\pm\) 1.09}  & 43.83 \tiny{\(\pm\) 0.34} &  \underline{56.09} \tiny{\(\pm\) 0.48} & 55.04 \tiny{\(\pm\) 1.07}& \underline{56.15} \tiny{\(\pm\) 0.63}\\
    H3K4me2&29.63 \tiny{\(\pm\) 1.68}& 32.38 \tiny{\(\pm\) 0.98}&  49.25 \tiny{\(\pm\) 2.06}& \underline{49.96} \tiny{\(\pm\) 0.90}& \textbf{55.59} \tiny{\(\pm\) 1.08}\\
    H3K4me3&38.76 \tiny{\(\pm\) 1.30} & 31.49 \tiny{\(\pm\) 3.40}  & 57.90 \tiny{\(\pm\) 0.92}& \underline{60.92} \tiny{\(\pm\) 0.72}& \textbf{63.68} \tiny{\(\pm\) 0.34}\\
    H3K79me3&60.98 \tiny{\(\pm\) 1.06}   & 60.48 \tiny{\(\pm\) 0.50} & \underline{71.94} \tiny{\(\pm\) 0.44} &70.97 \tiny{\(\pm\) 0.77}& \textbf{74.29} \tiny{\(\pm\) 0.08}\\
    H3K9ac&54.57 \tiny{\(\pm\) 0.71} & 52.55 \tiny{\(\pm\) 0.85} &\underline{64.35} \tiny{\(\pm\) 1.03}  &62.57 \tiny{\(\pm\) 0.46}& \textbf{64.78} \tiny{\(\pm\) 0.50}\\
    H4&79.19 \tiny{\(\pm\) 0.49} & 79.60 \tiny{\(\pm\) 0.55}  &\underline{80.71} \tiny{\(\pm\) 0.43}&78.73 \tiny{\(\pm\) 0.66}& \textbf{81.18} \tiny{\(\pm\) 0.25}\\
    H4ac&48.02 \tiny{\(\pm\) 1.40} &41.53 \tiny{\(\pm\) 0.24}  &63.03 \tiny{\(\pm\) 0.81} & \underline{63.06} \tiny{\(\pm\) 0.62}& \textbf{67.65} \tiny{\(\pm\) 0.39}\\
    \midrule
    Regulatory Annotation Avg. & 84.86 \tiny{\(\pm\) 0.01} & 84.83 \tiny{\(\pm\) 0.29} & \textbf{86.16} \tiny{\(\pm\) 0.17} & 84.87 \tiny{\(\pm\) 0.20}& \textbf{86.16} \tiny{\(\pm\) 0.07}\\
    \midrule
    Enhancer&78.16 \tiny{\(\pm\) 0.29} & 79.13 \tiny{\(\pm\) 0.56} & \underline{79.40} \tiny{\(\pm\) 0.29} & 78.81 \tiny{\(\pm\) 0.54}&  \textbf{79.90} \tiny{\(\pm\) 0.39}\\
    Enhancer Types&58.14 \tiny{\(\pm\) 1.02} & 54.73 \tiny{\(\pm\) 1.24} & \underline{59.84} \tiny{\(\pm\) 0.45} & 58.36 \tiny{\(\pm\) 0.58}& \textbf{60.50} \tiny{\(\pm\) 0.59}\\
    Promoter All&96.23 \tiny{\(\pm\) 0.61} &97.05 \tiny{\(\pm\) 0.05} & \textbf{97.37} \tiny{\(\pm\) 0.04} &  96.19 \tiny{\(\pm\) 0.14}& \underline{97.16} \tiny{\(\pm\) 0.12}\\
    Promoter Non-TATA&96.69 \tiny{\(\pm\) 0.16} & 97.02 \tiny{\(\pm\) 0.16}  & \textbf{97.65} \tiny{\(\pm\) 0.12} & 96.43 \tiny{\(\pm\) 0.09}& \underline{97.24} \tiny{\(\pm\) 0.14}\\
    Promoter TATA&95.10 \tiny{\(\pm\) 0.24} & \underline{96.22} \tiny{\(\pm\) 0.33} & \textbf{96.55} \tiny{\(\pm\) 0.24} &94.55 \tiny{\(\pm\) 0.38}& 96.01 \tiny{\(\pm\) 0.09}\\
    \midrule
    Splice Site Annotation Avg. & \textbf{98.71} \tiny{\(\pm\) 0.05} &98.02 \tiny{\(\pm\) 0.09} &96.69 \tiny{\(\pm\) 0.09} & 96.82 \tiny{\(\pm\) 0.11}& \underline{98.09} \tiny{\(\pm\)0.09}\\
    \midrule
    All&\textbf{98.36} \tiny{\(\pm\) 0.04} & 97.83 \tiny{\(\pm\) 0.06} & 95.68 \tiny{\(\pm\) 0.11} &96.78 \tiny{\(\pm\) 0.18}& \underline{98.14} \tiny{\(\pm\) 0.08}\\
    Accpetor&\textbf{98.69} \tiny{\(\pm\) 0.14} & 97.81 \tiny{\(\pm\) 0.28} & 97.71 \tiny{\(\pm\) 0.11} & 96.52 \tiny{\(\pm\)  0.19}& \underline{98.01} \tiny{\(\pm\)  0.13}\\
    Donor&\textbf{99.09} \tiny{\(\pm\) 0.05} & \underline{98.43} \tiny{\(\pm\) 0.05} &96.67 \tiny{\(\pm\) 0.17} &97.16 \tiny{\(\pm\) 0.16}& 98.10 \tiny{\(\pm\) 0.13}\\
    \bottomrule
  \end{tabular}}
\end{table}

\subsubsection{Nucleotide Transformer Benchmarks}

Next, we evaluate MxDNA on the Nucleotide Transformer Benchmarks~\cite{dalla2023nucleotide}, which includes 18 datasets across three task types: histone marker prediction, regulatory annotation prediction, and splice site annotation prediction. For this benchmark, the BERT-like models are finetuned for 20 epochs. Following~\cite{schiff2024caduceus}, we use the Matthews Correlation Coefficient (MCC) for histone markers tasks, F1 score for regulatory and splice site annotation tasks, except accuracy for splice site all task.

As shown in Table~\ref{performace NT}, MxDNA achieves the best performance on 10 out of 18 datasets and ranks in the top-2 on 16 out of 18 datasets. On average, MxDNA shows an improvement of 1.48 points compared to the second-best model, DNABERT2. Notably, MxDNA significantly outperforms all other models in the histone markers tasks while maintaining competitive performance in regulatory annotation and splice site annotation tasks.

\subsection{Ablation Studies}

\paragraph{Different Tokenization Methods:} We compare various tokenization methods by pretraining models for 100k steps using the same backbone but different tokenization methods. The results in Table~\ref{ablation1} show that our learnt tokenization significantly outperforms traditional methods such as non-overlapping K-mer, BPE, overlapping K-mer, and single nucleotide tokenization. Among the rule-based methods, single nucleotide tokenization performs best, possibly because it doesn't incorporate human biases and focuses solely on the raw data, though it may make it difficult to capture higher-level semantics. Conversely, non-overlapping K-mer might disrupt meaningful units, BPE might fail to segment DNA sequences meaningfully, and overlapping K-mer could suffer from information leakage.

\begin{table}
  \caption{Average results on Nucleotide Transformer Benchmarks and Genomic Benchmarks with different tokenization methods. We highlight the best values in \textbf{bold} type, \underline{underline} the second best.}
  \label{ablation1}
  \centering
  \begin{tabular}{lll}
    \toprule
    Method & NT Benchmarks & Genomic Benchmarks\\
    \midrule
    Single Nucleotide (1-mer)&  \underline{75.07} \tiny{\(\pm\) 0.26} & \underline{88.56} \tiny{\(\pm\) 0.02}  \\
    Overlapping k-mer (6-mer) & 74.35 \tiny{\(\pm\) 0.35} & 88.55 \tiny{\(\pm\) 0.07} \\
    Non-overlapping k-mer (6-mer) &  67.65 \tiny{\(\pm\) 0.17} & 86.83 \tiny{\(\pm\) 0.06} \\
    Byte-pair Encoding (4096 tokens)&  74.96 \tiny{\(\pm\) 0.16} & 87.30 \tiny{\(\pm\) 0.16} \\
    MxDNA (Learnt Tokenization) & \textbf{77.52} \tiny{\(\pm\) 0.18} & \textbf{88.89} \tiny{\(\pm\) 0.05} \\
    \bottomrule
  \end{tabular}
\end{table}

\paragraph{Different Components: } We assess the impact of individual components by incrementally integrating each into the baseline model and pretraining them for 100k steps. Starting with a baseline of single nucleotide tokenization, we sequentially add the Mixture of Convolution Experts, the deformable convolution and jitter noise, resulting in the proposed MxDNA. The results in Table~\ref{ablation2} show substantial performance gains from the Mixture of Convolution Experts alone, demonstrating the effectiveness of the idea which allows the model to learn tokenization autonomously rather than depending on predefined tokenization. There are also noticeable performance improvements contributed by the deformable convolution and jitter noise, showing the effectiveness of explicitly equipping the model with capabilities to handle discontinuities, overlaps and ambiguities. 

\begin{table}
  \caption{Average results on Nucleotide Transformer Benchmarks and Genomic Benchmarks with components added successively. We highlight the best values in \textbf{bold} type, \underline{underline} the second best.}
  \label{ablation2}
  \centering
  \begin{tabular}{lll}
    \toprule
    Method & NT Benchmarks & Genomic Benchmarks\\
    \midrule
    Single Nucleotide Baseline &  75.07 \tiny{\(\pm\) 0.26} & 88.56 \tiny{\(\pm\) 0.02} \\
    + Mixture of Convolution Experts & 77.00 \tiny{\(\pm\) 0.05} & 88.72 \tiny{\(\pm\) 0.07} \\
    + Deformable Convolution & \underline{77.35} \tiny{\(\pm\) 0.12} &  \underline{88.86} \tiny{\(\pm\) 0.18}\\
    + Jitter Noise (MxDNA)&  \textbf{77.52} \tiny{\(\pm\) 0.18} &  \textbf{88.89} \tiny{\(\pm\) 0.05} \\
    \bottomrule
  \end{tabular}
\end{table}

\subsection{Analysis}

We conduct an analysis of the tokenization behaviors of MxDNA against previous methods on both a sample and dataset level, and present a output embedding analysis at a token level. Notably, MxDNA exhibits unique tokenization strategy distinct from prior methods and is able to inherently capture and differentiate genomic functionalities at a token level during self-supervised pretraining, potentially providing new biological insights. Visualization details are in Appx~\ref{AppendixVis}.

\begin{figure}[ht]
  \centering
  \includegraphics[width=\textwidth]{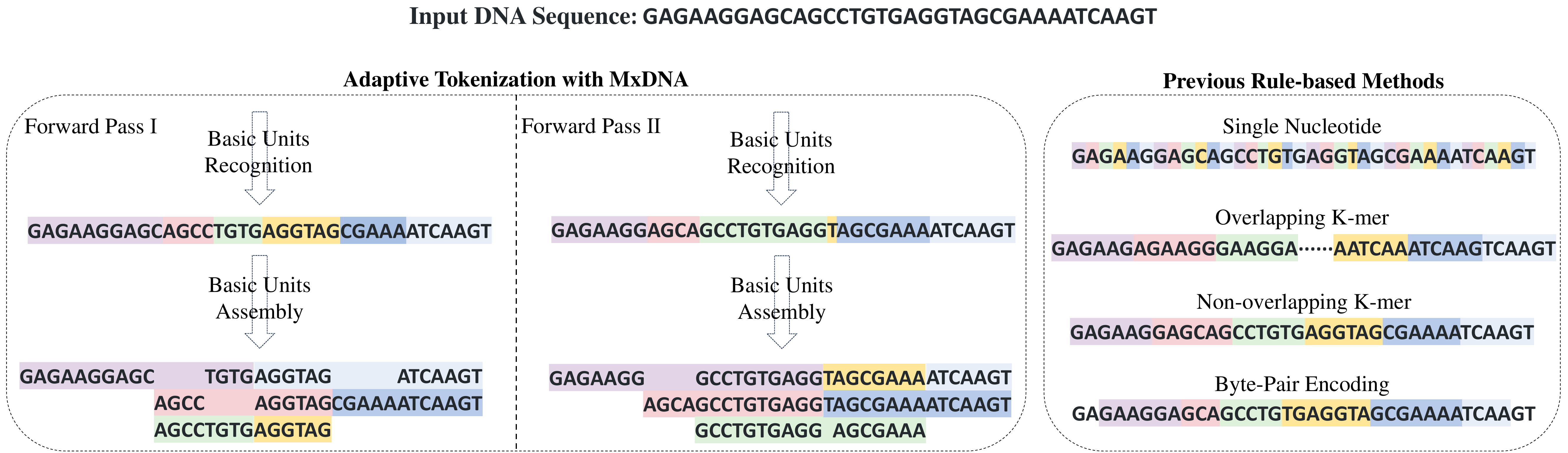}
  \caption{Tokenization results of MxDNA over two individual forward passes (left) compared to those of traditional rule-based methods (right). A block of the same colour refers to a single token.}
  \label{fig:tokenizationresult}
\end{figure}

\begin{figure}[ht]
  \centering
  \includegraphics[width=\textwidth]{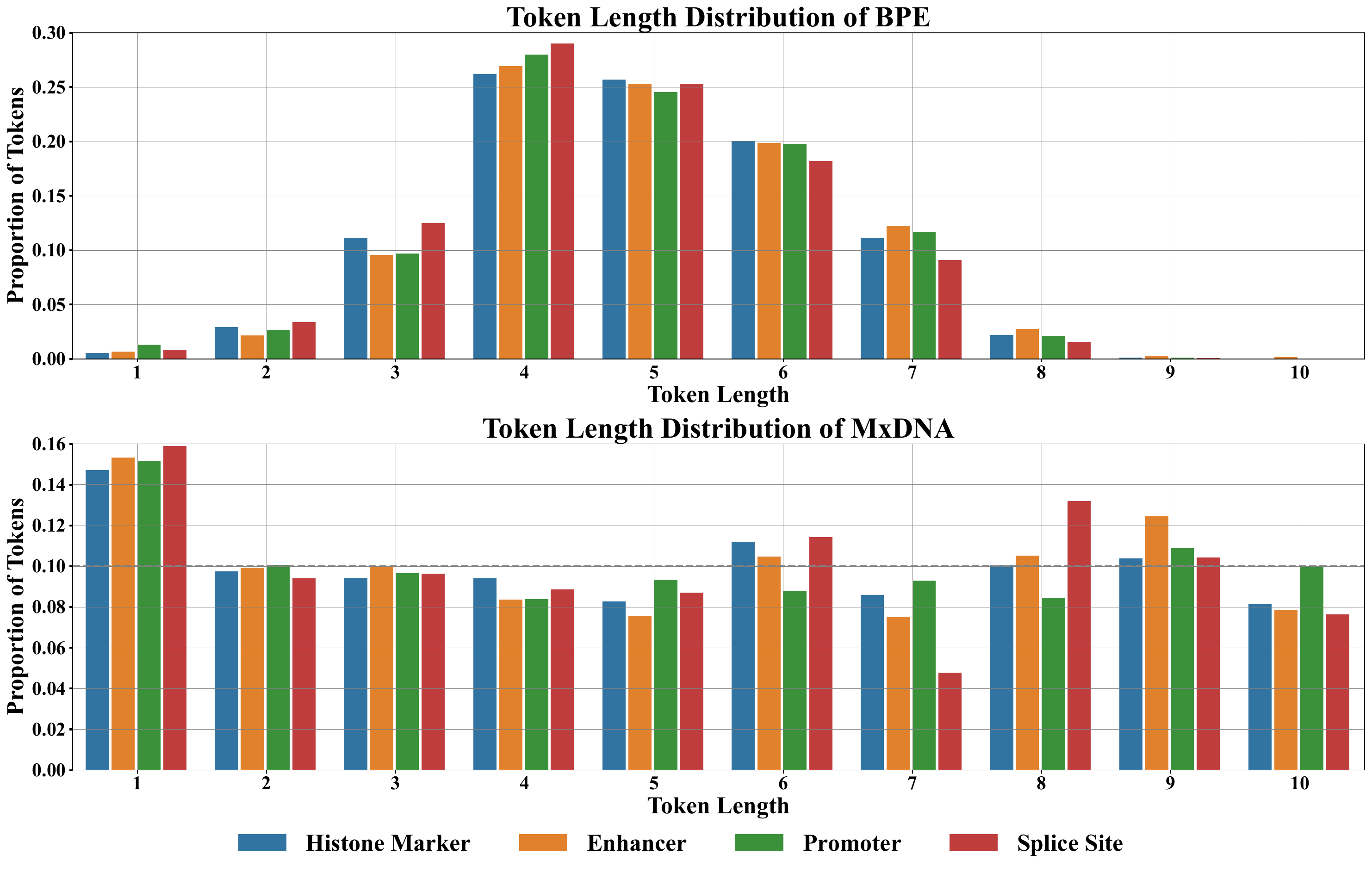}
  \caption{Distribution of token lengths for BPE (top) and MxDNA (bottom) across different downstream datasets, illustrating the distinct strategy of MxDNA for handling DNA tokenization. For the sake of simplicity, we regard the basic units as tokens for MxDNA.}
  \label{fig:tokendist}
\end{figure}

\paragraph{Sample Level} We first visualize the tokenization of a DNA sequence. For MxDNA, two individual forward passes with identical input and model yield slightly different results during training. It is worth mentioning that there are usually a small number of differences between the two results, and we display the region where the tokenization outcomes are different to show the ambiguous property for illustrative purposes. For previous rule-based methods, tokenization is static and performed only once. As depicted in Fig.~\ref{fig:tokenizationresult}, our learnt tokenization method tokenize the DNA sequence in a way distinctly different from previous rule-based method. Moreover, the discontinuous, overlapping and ambiguous tokenization results validate our design choices to manage these properties.

\begin{figure}[ht]

    \label{tokenembedding}
    \centering
    \includegraphics[width=\textwidth]{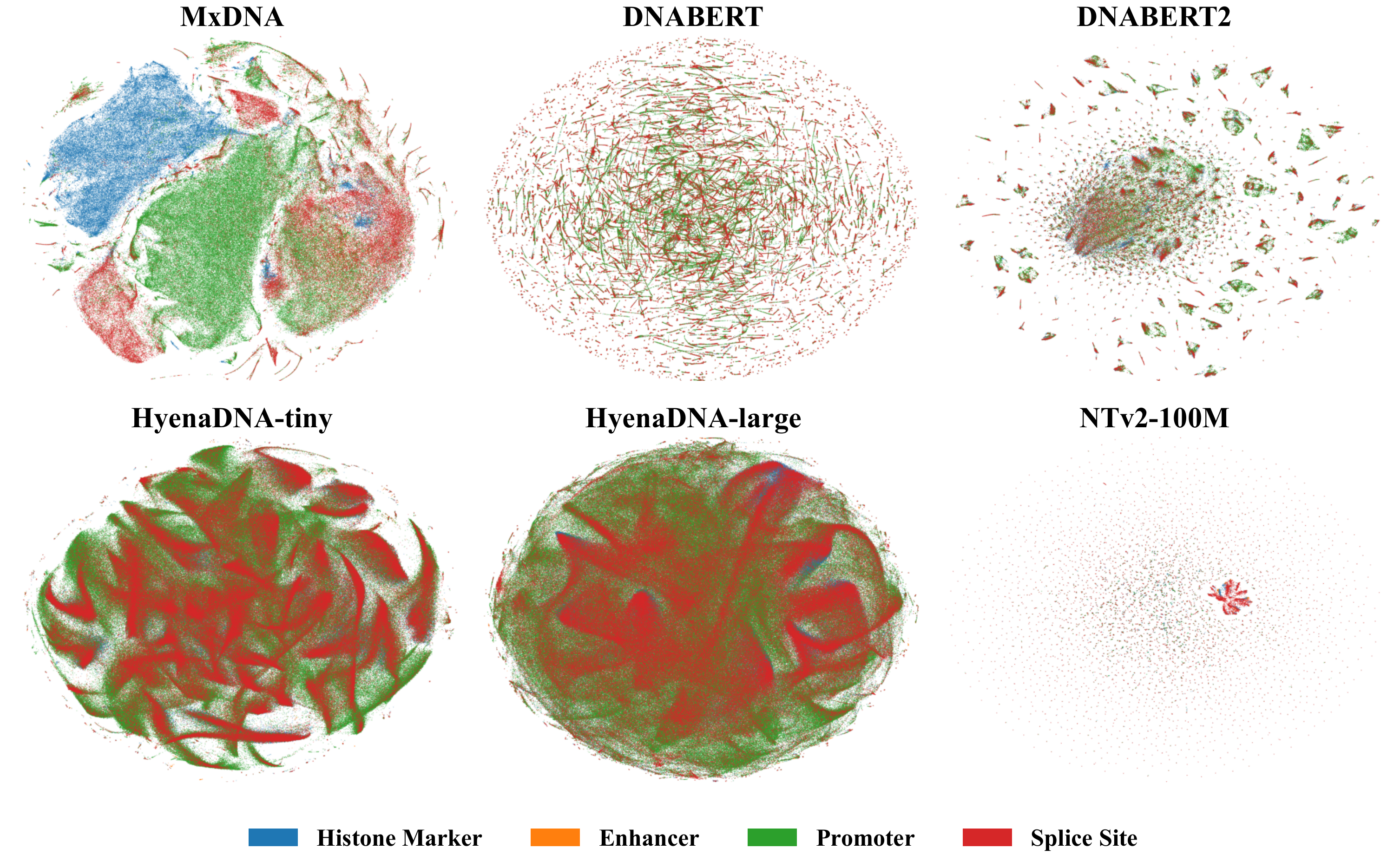}
    \caption{t-SNE visualization of the output embeddings at a token level across different functional sequences of different models, demonstrating MxDNA's unique capability to inherently capture and differentiate genomic functionalities at a token level.}
\end{figure}

\paragraph{Dataset Level} To gain more insights, we measure the distribution of token lengths across different downstream datasets for both MxDNA and BPE. For simplicity, we regard the basic units as tokens for MxDNA. BPE and MxDNA shows very distinct distribution of token lengths. As shown in Fig.~\ref{fig:tokendist}, BPE tends to produce a bell-shaped distribution, inherently biased by its frequency-based merging rule. Conversely, MxDNA's distribution is closer to a uniform distribution with preferences for specific lengths, reflecting its adaptive, task-oriented segmentation capabilities. Moreover, the variability in token distribution across datasets might suggest that DNA sequences of different functions might possess distinct patterns and meaningful units.

\paragraph{Token Embedding Analysis} Next, we use t-SNE to visualize the pretrained output tokens in sequences with different genomic functions of different foundation models. As is shown in Fig.~\ref{tokenembedding}, without any finetuning, the token embedding distributions of MxDNA are different across sequences with different functions: the tokens of Histone Marker, Promoter and Splice Site form unique clusters. While for all other foundation models, their tokens do not form clear clusters as MxDNA does. This shows MxDNA's superior capability to inherently capture and differentiate genomic functionalities at a token level, suggesting its robustness and specificity in representing biological sequences even before any supervised finetuning is applied.

\section{Conclusion}
\label{conclusion}

\paragraph{Summary} We present MxDNA, a framework developed to autonomously learn effective DNA tokenization strategies solely through gradient descent. MxDNA demonstrates strong performance against existing sota models and tokenization methods across 26 diverse genomic tasks in Nucleotide Transformer Benchmarks and Genomic Benchmarks with no additional cost. We also perform an analysis of the tokenization mechanism and the token embedding space of MxDNA, showing its distinct tokenization strategy against previous methods and unique capability to capture genomic functionalities at a token level.

\paragraph{Limitations \& Future Works} While MxDNA demonstrates strong quantitative performance on various downstream tasks, direct biological validation of the model's tokenization decision remains limited. Furthermore, the evaluation on long range tasks is lacking due to quadratic cost of self-attention, although the learnt tokenization is expected to help reduce sequence length effectively and can be combined with sub-quadratic architectures~\cite{gu2023mamba,poli2023hyena}. Future research will focus on refining MxDNA's design to learn a better and more interpretable tokenization strategy, and testing its applicability to broader genomic analyses especially on more long range tasks.

{
\small
\bibliography{neurips_2024}
\bibliographystyle{unsrt}
}

\newpage
\appendix
\section{Appendix / supplemental material}

\subsection{Extended Related Works}

\label{AppendixRelated}

\subsubsection{Image Tokenization}
Tokenization in computer vision (CV) attempts to transform images into formats that can be efficiently processed by machine learning models especially transformers. In line with Character-level tokenization,~\cite{chen2020generative} directly using raw pixels as input units. The Vision Transformer (ViT)~\cite{dosovitskiy2020image} splits an image into patches of identical sizes and treats these patches as tokens in NLP, demonstrating remarkable performance on standard image recognition tasks. Inspired by ~\cite{van2017neural},  ~\cite{ramesh2021zero,bao2021beit} utilized an image tokenizer learnt by discrete variational autoencoder (dVAE) to map pixels into discrete tokens according to a visual codebook. Meanwhile, researchers also utilize detection or segmentation features for visual representations. For instance, ~\cite{anderson2018bottom, bugliarello2021multimodal} used a pretrained Faster R-CNN model~\cite{ren2015faster} to extract region features. Recently, ~\cite{chen2024subobject, kim2024vision} exploit Segment Anything Model (SAM)~\cite{kirillov2023segment} to construct a sub-word tokenization and semantic tokenization respectively.

\subsubsection{Mixture of Experts} 

(Sparse) MoE is first designed to improve the capacity of neural networks while maintaining total computations. ~\cite{shazeer2017outrageously} uses MoE as a general purpose neural network component and realizes sparse gating, demonstrating its use as a practical way to massively increase model capacity. By replacing FNN with Mixture of Experts, \cite{fedus2022switch, jiang2024mixtral} successfully combine sparse MoE and Transformers, achieving superior capabilities with less computational cost. Previous methods generally replace a layer of the neural networks with multiple, sparsely activated identical alternatives, governed by a gating mechanism. Recently, ~\cite{chalapathi2024scaling} explicitly adds interpretability to each expert by letting each expert solve the constraint over smaller decomposed domains through differentiable optimization.

\subsubsection{Deformable Convolution}

Deformable convolution explicitly equips the model with ability to adapt to the geometric variations of different objects~\cite{dai2017deformable, zhu2019deformable}. Unlike the attention mechanism, which focuses on capturing long-range relationships, deformable convolution locally samples feature maps using learnt offsets and modulation factors. By modeling complex geometric transformations effectively, deformable convolution networks achieve significant performance improvements in various tasks, including image classification, object detection and semantic segmentation.

\subsubsection{Combination of Convolution and Transformer} 

The integration of convolutional layers with transformer architectures has emerged as a powerful approach across various domains, effectively combining the strengths of both techniques. Conformer~\cite{gulati2020conformer} applies this hybrid design to audio processing, enhancing the capture of local and global dependencies in audio signals. In computer vision, the CvT~\cite{wu2021cvt} introduces convolutions into transformers to improve efficiency and representational power, while Early Convolution~\cite{xiao2021early} in Vision Transformer incorporates convolutional layers early in the architecture to enhance input feature representations. Extending to genomic data, Enformer~\cite{avsec2021effective} applies a similar approach as in~\cite{xiao2021early} to model complex dependencies in DNA sequences and reduce the computational cost, showcasing the potential of hybrid architectures to handle highly specialized data types like genomic sequences.

\begin{table}
\centering
\begin{tabular}{lp{10cm}}
\hline
\textbf{Term} & \textbf{Description} \\ \hline
$l$ & Number of nucleotides \\
$d$ & Dimension of hidden states \\
$n$ & Number of experts \\
$k$ & Number of basic units \\
$f$ & Kernel size of the deformable convolution\\
$i$ & Indices of nucleotides or tokens \\
$j$ & Indices of experts \\
$\mathbf{X} \in \mathbb{R}^{l \times d}$ & Input nucleotide sequence \\
$\mathbf{S} \in \mathbb{R}^{l \times n}$ & Confidence scores of basic units existence \\
$\mathbf{L} \in \mathbb{N}^{n}$ & Kernel sizes of convolution experts \\
$\mathbf{M} \in \mathbb{N}^{l}$ & Mask indicating the existence of basic units \\
$\mathbf{E_j} \in \mathbb{R}^{L_j \times d} \rightarrow \mathbb{R}^d$ & Convolution experts \\
$\mathbf{U} \in \mathbb{R}^{k \times d}$ & Basic units \\
$\Delta \mathbf{P} \in \mathbb{R}^{k \times f}$ & Offsets ofthe deformable convolution \\
$\Delta \mathbf{M} \in \mathbb{R}^{k \times f}$ & Modulation factors of the deformable convolution  \\
$\mathbf{T} \in \mathbb{R}^{k \times d}$ & Final tokens \\ \hline
\end{tabular}
\caption{Glossary of terms used in describing the method.}
\label{tab:glossary}
\end{table}

\subsection{Method Details}
\label{AppendixA2}

For the convolution expert, we adapt design principles from~\cite{gulati2020conformer}. Our 1-D convolution expert starts with a pointwise convolution \(W^{in}_{pj}\) paired with a Gated Linear Unit (GLU), followed by a 1-D grouped convolution \(W_{gj}\). Subsequent to the grouped convolution, a Layer Normalization (LayerNorm) and a Swish activation layer \(W^{out}_{pj}(\text{Swish}())\) are applied. The grouped convolution here has number of groups equal to the factor of hidden size closest to kernel size, and number of output channel equal to number of input channels. This ensures that each convolution expert has similar parameter counts in spite of different kernel sizes.

\begin{equation}
E_j\left(X) = W^{out}_{pj} \left(\text{Swish}\left(\text{LayerNorm}(W_{gj} \ast \text{GLU}\left(W^{in}_{pj}  X\right)\right)\right)\right)
\end{equation}

\subsubsection{Non-Maximum Suppression}

The pseudocode in~\ref{alg1} describes the selection process for optimal basic units based on scores, ensuring no overlaps, and using kernel sizes to guide the selection.

The input consists of: positions (all nucleotide positions), kernel sizes (all kernel sizes), scores (scores for each (position, kernel size) pair) for the possibility of a basic unit of a given size existing at a given position. The output is a mask indicating selected basic units with their corresponding kernel sizes.

\begin{algorithm}
\caption{Detailed Non-Maximum Suppression for Basic Unit Placement}
\begin{algorithmic}[1] 
\Procedure{NMS}{positions $P = [1, 2, \ldots, l]$, kernel sizes $L \in \mathbb{N}^n$, scores $S \in \mathbb{R}^{l \times n}$}
    \State Sort all ($P_i$, $L_j$) pairs by $S_{ij}$ in descending order, where $i \in [1, 2, \ldots, l]$ and $j \in [1, 2, \ldots, n]$.
    \State Initialize an output array with zeros $M \in \mathbb{N}^l$.
    \For{each ($P_i$, $L_j$) pair in the sorted pairs}
        \State Calculate the start and end of the region at $P_i$ with width $L_j$.
        \If{the region is not overlapped with any region in $M$}
            \State $M_{P_i} \in \mathbb{N} \leftarrow L_j$
        \EndIf
    \EndFor
    \State \Return $M$.
\EndProcedure
\end{algorithmic}
\label{alg1}
\end{algorithm}

\subsubsection{Sparse Mixture of Convolution Experts}

The pseudocode in~\ref{alg2} outlines the selective activation of convolutions at positions determined by Non-Maximum Suppression, using corresponding kernel sizes.

The input consists of: input (embeddings of nucleotides), positions (at a nucleotide level) of selected basic units with their corresponding kernel sizes. The output is the embeddings of the selected basic units.

\begin{algorithm}
\caption{Detailed Sparse Convolution}
\begin{algorithmic}[1]
\Procedure{Sparse Convolution}{input $X \in \mathbb{R}^{l \times d}$, selected positions with kernel sizes $M \in \mathbb{N}^l$}
    \State $k$ $\leftarrow$ number of non-zero elements in $M$.
    \State Initialize an output array with zeros $U \in \mathbb{R}^{k \times d}$.
    \State Initialize counter $cnt = 0$.
    \For{each $i$ in $[1, 2, \ldots, l]$}
        \If{$M_i \neq 0$}
            \State cnt $\leftarrow$ cnt + 1.
            \State Extract the segment of $X$ centered at $i$ with width $M_i$.
            \State $U_{cnt} \in \mathbb{R}^d$ $\leftarrow$ Apply the convolution expert of kernel size $M_{i}$ to the segment.
        \EndIf
    \EndFor
    \State \Return $U$.
\EndProcedure
\end{algorithmic}
\label{alg2}
\end{algorithm}

\subsubsection{Deformable Convolution}

The pseudocode in~\ref{alg3} details how deformable convolution dynamically adjusts based on input features by modifying its parameters for each input segment.

The input consists of: input (embeddings of selected basic units). The output is the embeddings of the final tokens.

\begin{algorithm}
\caption{Detailed Deformable Convolution}
\begin{algorithmic}[1]
\Procedure{Deformable Convolution}{input $U \in  \mathbb{R}^{k \times d}$}
    \State Initialize an output array with zeros $T \in \mathbb{R}^{k \times d}$.
    \For{each $i$ in $[1, 2, \ldots, k]$}
        \State Calculate offsets $\Delta P_i \in \mathbb{R}^f$ based on $U_i$.
        \State Calculate modulation factors $\Delta M_i \in \mathbb{R}^f$ based on $U_i$.
        \State Extract the deformed segment of $U$ centred at $i$ according to $\Delta P_i$.
        \State Weight the segment by $\Delta M_i$.
        \State $T_i \in \mathbb{R}^d$ $\leftarrow$ Compute the segment's dot product with the convolution kernel of size $f$.
    \EndFor
    \State \Return $T$.
\EndProcedure
\end{algorithmic}
\label{alg3}
\end{algorithm}

\subsection{Back Propagation}
\label{AppendixA3}
The tokenization module is learnt solely through gradient descent. In this section, we will focus on the gradients with respect to the Basic Units Scoring Block and Distal Relation Estimation Block.

\subsubsection{Sparse Mixture of Convolution Experts}

Recall the forward process in Eq.~\ref{SMoCEeq}:

\begin{equation}
\begin{aligned}
&U_i = E_j\left(X_{\left[i-\left \lceil\frac{M_i}{2}\right \rceil+1:i+\left \lfloor\frac{M_i}{2}\right \rfloor\right]}\right) \cdot \text{softmax}\left(S_i\right)_{j} \cdot \boldone\left(M_i > 0\right), \text{where } L_j = M_i \\
&\boldone(M_i > 0)=
\begin{cases}
    1, M_i > 0 \\
    0, M_i \leq 0
\end{cases} ,
\text{softmax}(S_i)_{j} = \frac{e^{S_{ij}}}{\sum_{k} e^{S_{ik}}} ,
\end{aligned}
\end{equation}

The gradient w.r.t the score \(S_{ik}\) is as follows:

\begin{equation}
    \begin{aligned}
        & \frac{\partial U_i}{\partial S_{ik}} =  E_j\left(X_{\left[i-\left \lceil\frac{M_i}{2}\right \rceil+1:i+\left \lfloor\frac{M_i}{2}\right \rfloor\right]}\right) \cdot \boldone(M_i > 0) \cdot \frac{\partial \text{softmax}(S_i)_j}{\partial S_{ik}}, \text{where } L_j = M_i \\
        & \frac{\partial \text{softmax}(S_i)_j}{\partial S_{ik}} = 
        \begin{cases}
            \text{softmax}(S_i)_j \cdot \left(1-\text{softmax}(S_i)_j\right) &, k = j \\
            -\text{softmax}(S_i)_j \cdot \text{softmax}(S_i)_k &, k \neq j
        \end{cases}
    \end{aligned}
\end{equation}

\subsubsection{Deformable Convolution}

Recall the forward process in Eq.~\ref{DCNeq}:  

\begin{equation}
T_i = \sum_{p \in \left\{-\left \lceil\frac{f}{2}\right \rceil + 1,...,\left \lfloor\frac{f}{2}\right \rfloor\right\}} w_p \cdot U_{i+p+\Delta p} \cdot \Delta m
\end{equation}

\begin{equation}
U_{p'} = \sum_{q \in \left\{1,...,k\right\}} \max\left(0, 1 - \left|p' - q\right|\right) \cdot U_q
\end{equation}

The gradients w.r.t the offset \(\Delta p\) and the modulation \(\Delta m\) are as follows:

\begin{equation}
    \frac{\partial T_i}{ \partial \Delta p} = \sum_{p \in \left\{-\left \lceil\frac{f}{2}\right \rceil + 1,...,\left \lfloor\frac{f}{2}\right \rfloor\right\}} w_p \cdot \Delta m \cdot \sum_{q \in \left\{1,...,k\right\}} \frac{\partial \max\left(0, 1 - \left|i+p+\Delta p - q\right|\right) }{\Delta p}\cdot U_{q} 
\end{equation}

\begin{equation}
\begin{aligned}
    \frac{\partial T_i}{ \partial \Delta m} = \sum_{p \in \left\{-\left \lceil\frac{f}{2}\right \rceil + 1,...,\left \lfloor\frac{f}{2}\right \rfloor\right\}} w_p \cdot U_{i+p+\Delta p} \cdot \frac{\partial \Delta m}{\partial \Delta m} 
    = \sum_{p \in \left\{-\left \lceil\frac{f}{2}\right \rceil + 1,...,\left \lfloor\frac{f}{2}\right \rfloor\right\}} w_p \cdot U_{i+p+\Delta p}
\end{aligned}
\end{equation}

\subsection{Experiment Setting Details}
\label{AppendixExp}
\subsubsection{Settings}

\paragraph{Model Implementation} MxDNA is built on the Nucleotide Transformer V2 architecture which incorporates several architectural improvements recognized in the NLP community, such as rotary positional encodings \cite{su2024roformer}, SwishGLU MLP \cite{shazeer2020glu}, and the exclusion of linear bias terms \cite{touvron2023llama, chowdhery2023palm}. Consistent with Nucleotide Transformer V2 100M, MxDNA has 512 hidden units, an expansion factor of 4, 16 attention heads, and 22 layers, totaling approximately 100M parameters. Specifically, the model's learnt tokenization module includes 10 convolution experts with kernel sizes ranging from 1 to 10, along with a deformable convolution block with a kernel size of three. We integrate this module by replacing the fifth transformer block, aiming to avoid introducing additional computations. We utilize FlashAttention~\cite{dao2022flashattention,dao2023flashattention} for efficient attention calculations.

\paragraph{Pretraining} Following \cite{ji2021dnabert}, MxDNA is pretrained on the whole Human Reference Genome \cite{HG38} using Masked Language Modeling. We removed all sequences gaps and unannotated regions and extracted 70 to 510-nt-long sequences as training data. We mask 15\% of the tokens, with 80\% replaced by a special [MASK] token, 10\% replaced with a random vocabulary token, and 10\% left unchanged. All masking happens at the initial input stage(single nucleotide, 6mer tokens, bpe tokens). For model using single nucleotide tokenization, non-overlapping 6mer and BPE, the masking is performed randomly and mask out 15\% of total tokens except of special tokens. For model using overlapping 6mer, we follow the strategy used in \cite{ji2021dnabert}, with contiguous k-length spans of certain k-mers are masked, totalling around 15\% of the tokens. An auxiliary balancing loss with a weight of 0.01 is used to prevent degradation towards a single expert, following \cite{fedus2022switch}. The model is trained with a learning rate of 1e-4 and a batch size of 512. We employ the AdamW optimizer with \(\beta_1 = 0.9\), \(\beta_2 = 0.98\), \(\epsilon = 1e-6\), a weight decay of 0.01, and a cosine annealing learning rate scheduler with a linear warm-up over the first 10\% of steps. The model undergoes training for 500k steps for main performance comparisons and 100k steps for ablations.

\paragraph{Downstream} We download the data from \url{https://huggingface.co/spaces/InstaDeepAI/nucleotide_transformer_benchmark} for Nucleotide Transformer Benchmarks and \url{https://github.com/ML-Bioinfo-CEITEC/genomic_benchmarks} for Genomic Benchmarks. Moreover, in Nucleotide Transformer Benchmarks, the BERT-like models are finetuned using PEFT (parameter efficient finetuning) without providing the exact hyperparameters. Believing that fully fine-tuning these models will better leverage their capabilities and provide a fairer comparison , we decide to proceed with full finetuning for all models. We keep the original data splits in~\cite{dalla2023nucleotide,grevsova2023genomic}. We do not perform cross validation as~\cite{schiff2024caduceus} does since it will be too computationally expensive for BERT-like models and we decide to follow the practice of HyenaDNA~\cite{nguyen2024hyenadna} instead. Additionally, we repeat all experiments under three random seeds, report the average results with sample standard deviations.

All the BERT-like models are fully finetuned with a batch size of 32 and a learning rate of 3e-5. We employ the AdamW optimizer with \(\beta_1 = 0.9\), \(\beta_2 = 0.999\), \(\epsilon = 1e-8\), and a weight decay of 0.01. Models are trained for 10 epochs on Genomic Benchmarks and 20 epochs on Nucleotide Transformer Benchmarks, with the learning rate linearly warmed up over the first epoch and then decaying to zero during the remaining epochs. For the Mouse Enhancers dataset (sequence lengths with mean = 2381, std = 984.4, max = 4707), we truncate the sequence to a maximum length of 4096, which is considered acceptable. For DNABERT, which can not handle sequences of length over 512, we truncate the sequence to a maximum length of 512. 

For HyenaDNA, we fully finetune the pretrained model from \url{https://huggingface.co/LongSafari/hyenadna-tiny-1k-seqlen-d256} using the hyperparameters provided by~\cite{nguyen2024hyenadna} in docker image \url{hyenadna/hyena-dna-nt6:latest} for Nucleotide Transformer Benchmarks, and \url{https://huggingface.co/LongSafari/hyenadna-tiny-1k-seqlen} with modified hyperparameters recommended by~\cite{schiff2024caduceus} for Genomic Benchmarks. Their research suggests that training with sequence lengths 2 to 4 times the length of sequences used in downstream tasks typically yields the best performance. Thus, the tiny models are the best choice for most of the downstream tasks in Nucleotide Transformer Benchmarks and Genomic Benchmarks since most of tasks have sequence length of around a few hundreds and the tiny model are pretrained with 1000 length sequence. Notice that although our reproduced results is a bit lower than the results reported by the authors of HyenaDNA, the performance of MxDNA is still better than originally reported results on most of the tasks.

\subsubsection{Metrics}
This section defines the metrics used to evaluate the performance of models on various genomic tasks. On Nucleotide Transformer Benchmarks, We used the Matthews Correlation Coefficient (MCC) for histone marker tasks, F1 scores for regulatory and splice site annotation tasks, except accuracy for splice site all task. Top-1 Accuracy is used for all tasks in Genomics Benchmarks.

\paragraph{Matthews Correlation Coefficient (MCC)}
The Matthews Correlation Coefficient is a robust statistical rate which takes into account true and false positives and negatives and is generally regarded as a balanced measure that can be used even if the classes are of very different sizes.

\[
\text{MCC} = \frac{TP \times TN - FP \times FN}{\sqrt{(TP+FP)(TP+FN)(TN+FP)(TN+FN)}}
\]
where \( TP \), \( TN \), \( FP \), and \( FN \) are the numbers of true positives, true negatives, false positives, and false negatives, respectively.

\paragraph{F1 Score}
We use the macro-averaged F1 score, which is computed using the arithmetic mean of  all the per-class F1 scores. The (per-class) F1 score is the harmonic mean of precision and recall and is particularly useful when the costs of false positives and false negatives are high. 

\[
\text{(per-class) F1} = 2 \times \frac{\text{precision} \times \text{recall}}{\text{precision} + \text{recall}}
\]
where \( \text{precision} = \frac{TP}{TP + FP} \) and \( \text{recall} = \frac{TP}{TP + FN} \).

\paragraph{Accuracy}
Accuracy is the proportion of true results (both true positives and true negatives) among the total number of cases examined.

\[
\text{Accuracy} = \frac{TP + TN}{TP + TN + FP + FN}
\]

\subsection{Ablation Results Details}

\begin{table}
  \caption{Genomic Benchmarks. Different tokenization methods.}
  \label{AblationFullGBTok}
  \centering
  \scalebox{0.9}{
  \begin{tabular}{llllll}
    \toprule
    Dataset   &1-mer&Ovlp 6-mer&Non-Ovlp 6-mer&BPE & MxDNA \\
    \midrule
    Average & 88.56 \tiny{\(\pm\) 0.02} & 88.55 \tiny{\(\pm\) 0.07} & 86.83 \tiny{\(\pm\) 0.06} &87.30  \tiny{\(\pm\) 0.16}& 88.89  \tiny{\(\pm\) 0.05}\\
    \midrule
    Mouse Enhancers& 77.56 \tiny{\(\pm\) 0.94} &78.81 \tiny{\(\pm\) 0.85}  & 82.43 \tiny{\(\pm\) 0.90} &80.44  \tiny{\(\pm\) 1.44}&   80.44  \tiny{\(\pm\) 0.63}\\
    Coding vs Intergenomic&95.05 \tiny{\(\pm\) 0.13} &94.97 \tiny{\(\pm\) 0.06} &92.73 \tiny{\(\pm\) 0.08} &92.25  \tiny{\(\pm\) 0.11}&94.78  \tiny{\(\pm\) 0.04}\\
    Human vs Worm&97.52 \tiny{\(\pm\) 0.04}&97.14 \tiny{\(\pm\) 0.11}& 96.35 \tiny{\(\pm\) 0.04}  &96.59  \tiny{\(\pm\) 0.01}&97.27  \tiny{\(\pm\) 0.07}\\
    Human Enhancer Cohn&73.70 \tiny{\(\pm\) 0.53} &  73.07 \tiny{\(\pm\) 0.56}  &72.72 \tiny{\(\pm\) 0.19} & 72.92  \tiny{\(\pm\) 0.23}& 73.98  \tiny{\(\pm\) 0.40}\\
    Human Enhancer Ensembl&92.79 \tiny{\(\pm\) 0.09}& 92.98 \tiny{\(\pm\) 0.07}  & 91.79 \tiny{\(\pm\) 0.23} &92.38  \tiny{\(\pm\) 0.08}& 92.73  \tiny{\(\pm\) 0.08}\\
    Human Regulatory&94.03 \tiny{\(\pm\) 0.02} & 94.21 \tiny{\(\pm\) 0.08} & 93.73 \tiny{\(\pm\) 0.08}&89.70  \tiny{\(\pm\) 0.10}& 94.10  \tiny{\(\pm\) 0.12}\\
    Human OCR Ensembl&80.84 \tiny{\(\pm\) 0.50} & 81.36 \tiny{\(\pm\) 1.05}  &75.64 \tiny{\(\pm\) 0.64} &77.87  \tiny{\(\pm\) 0.45}&80.62  \tiny{\(\pm\) 0.42}\\
    Human NonTATA Promoters&97.00 \tiny{\(\pm\) 0.05} &95.84 \tiny{\(\pm\) 0.69} & 89.24 \tiny{\(\pm\) 0.24}  & 96.22  \tiny{\(\pm\) 0.17}&97.22  \tiny{\(\pm\) 0.23}\\
    \bottomrule
  \end{tabular}}
\end{table}

\begin{table}
  \caption{Nucleotide Transformer Benchmarks. Different tokenization methods.}
  \label{AblationFullNTTok}
  \centering
  \scalebox{0.93}{
  \begin{tabular}{llllll}
    \toprule
    Dataset   &1-mer&Ovlp 6-mer&Non-Ovlp 6-mer&BPE & MxDNA \\
    \midrule
    Average &75.07 \tiny{\(\pm\) 0.26} & 74.35 \tiny{\(\pm\) 0.35} & 67.65 \tiny{\(\pm\) 0.17} &74.96 \tiny{\(\pm\) 0.16}& 77.52 \tiny{\(\pm\) 0.18} \\
    \midrule
    Histone Markers Avg. & 63.13 \tiny{\(\pm\) 0.34} & 61.88 \tiny{\(\pm\) 0.66} & 50.36 \tiny{\(\pm\) 0.28} & 64.58  \tiny{\(\pm\) 0.13} & 67.29 \tiny{\(\pm\) 0.23} \\
    \midrule
    H3& 80.92 \tiny{\(\pm\) 0.85} &80.94 \tiny{\(\pm\) 0.64}  & 74.77 \tiny{\(\pm\) 0.32} & 80.26 \tiny{\(\pm\) 0.15} & 82.14 \tiny{\(\pm\) 0.76}\\
    H3K14ac&62.00 \tiny{\(\pm\) 1.79}& 62.17 \tiny{\(\pm\)0.80} &46.33 \tiny{\(\pm\) 0.72} & 65.91 \tiny{\(\pm\) 0.72} & 68.29 \tiny{\(\pm\)0.65}\\
    H3K36me3&62.59 \tiny{\(\pm\) 1.50} &63.26 \tiny{\(\pm\) 1.60}& 50.49 \tiny{\(\pm\) 1.72} &63.83 \tiny{\(\pm\) 0.73} & 65.46 \tiny{\(\pm\) 1.74}\\
    H3K4me1&51.66\tiny{\(\pm\) 0.57}  & 50.11 \tiny{\(\pm\) 3.63} &  41.26 \tiny{\(\pm\) 1.05} & 54.33 \tiny{\(\pm\) 0.53} & 54.97 \tiny{\(\pm\) 1.50}\\
    H3K4me2&49.51 \tiny{\(\pm\) 0.95}& 35.59 \tiny{\(\pm\) 5.72}&  29.99 \tiny{\(\pm\) 0.93}& 49.48 \tiny{\(\pm\) 0.74} &  55.30 \tiny{\(\pm\) 0.49}\\
    H3K4me3&54.14 \tiny{\(\pm\) 0.95}& 54.15 \tiny{\(\pm\) 0.56}  & 30.83 \tiny{\(\pm\) 0.59}& 58.10 \tiny{\(\pm\) 0.55} & 63.82 \tiny{\(\pm\) 0.92}\\
    H3K79me3&70.36\tiny{\(\pm\) 1.58}   & 71.30 \tiny{\(\pm\) 0.04} & 59.99 \tiny{\(\pm\) 0.45} & 70.89 \tiny{\(\pm\) 0.74} & 73.74 \tiny{\(\pm\) 0.78}\\
    H3K9ac&60.63 \tiny{\(\pm\) 2.73} & 64.70 \tiny{\(\pm\) 0.32} &50.24 \tiny{\(\pm\) 1.31}  & 62.22 \tiny{\(\pm\) 0.54} & 63.15 \tiny{\(\pm\) 0.26}\\
    H4&80.23 \tiny{\(\pm\) 0.79} & 80.09 \tiny{\(\pm\) 0.59}  &78.27 \tiny{\(\pm\) 0.60}& 79.29 \tiny{\(\pm\) 0.41} & 80.89 \tiny{\(\pm\) 0.23}\\
    H4ac&59.25 \tiny{\(\pm\) 1.36} &56.49 \tiny{\(\pm\) 1.02}  &41.42 \tiny{\(\pm\) 0.89} & 61.45 \tiny{\(\pm\) 0.80}& 65.14 \tiny{\(\pm\) 0.23}\\
    \midrule
    Regulatory Annotation Avg. & 85.38 \tiny{\(\pm\) 0.15} & 84.95 \tiny{\(\pm\) 0.12} & 84.13 \tiny{\(\pm\) 0.09} & 84.62  \tiny{\(\pm\) 0.32}& 85.70 \tiny{\(\pm\) 0.21}\\
    \midrule
    Enhancer&78.93 \tiny{\(\pm\) 0.70} & 76.65 \tiny{\(\pm\) 0.80} & 78.48 \tiny{\(\pm\) 0.88} & 77.17 \tiny{\(\pm\) 0.80} &  79.73 \tiny{\(\pm\) 0.42}\\
    Enhancer Types&58.90 \tiny{\(\pm\) 0.72} & 59.12 \tiny{\(\pm\) 0.13} & 56.50 \tiny{\(\pm\) 0.61} & 59.57 \tiny{\(\pm\) 0.39}& 59.79 \tiny{\(\pm\) 0.52}\\
    Promoter All&96.92 \tiny{\(\pm\) 0.08} &96.94 \tiny{\(\pm\) 0.12} & 95.76 \tiny{\(\pm\) 0.10} &  95.77 \tiny{\(\pm\) 0.09} & 96.87 \tiny{\(\pm\) 0.10}\\
    Promoter Non-TATA&97.04 \tiny{\(\pm\) 0.05} & 96.84 \tiny{\(\pm\) 0.07}  & 95.87 \tiny{\(\pm\) 0.08} & 95.83  \tiny{\(\pm\) 0.16} & 96.81 \tiny{\(\pm\) 0.13}\\
    Promoter TATA&95.09 \tiny{\(\pm\) 0.34} & 95.21 \tiny{\(\pm\) 0.18} & 94.02 \tiny{\(\pm\) 0.01} & 94.77  \tiny{\(\pm\) 0.33} & 95.32 \tiny{\(\pm\) 0.17}\\
    \midrule
    Splice Site Annotation Avg. & 97.70 \tiny{\(\pm\) 0.19} &98.25 \tiny{\(\pm\) 0.06} &97.85 \tiny{\(\pm\) 0.08} & 93.46  \tiny{\(\pm\) 0.02} & 98.00 \tiny{\(\pm\)0.13}\\
    \midrule
    All&98.07 \tiny{\(\pm\) 0.17} & 98.18 \tiny{\(\pm\) 0.11} & 97.91 \tiny{\(\pm\) 0.15} &93.25 \tiny{\(\pm\) 0.41} &98.05 \tiny{\(\pm\) 0.13}\\
    Accpetor&97.61 \tiny{\(\pm\) 0.44} & 98.18 \tiny{\(\pm\) 0.15} & 97.65 \tiny{\(\pm\) 0.15} & 94.40  \tiny{\(\pm\) 0.34} & 97.68 \tiny{\(\pm\) 0.20}\\
    Donor&97.42 \tiny{\(\pm\) 0.58} & 98.39 \tiny{\(\pm\) 0.16} &98.00 \tiny{\(\pm\) 0.16} & 92.72  \tiny{\(\pm\) 0.42} & 98.28 \tiny{\(\pm\) 0.07}\\
    \bottomrule
  \end{tabular}}
\end{table}

\begin{table}
  \caption{Genomic Benchmarks. Different components.}
  \label{AblationFullGBCom}
  \centering
  \scalebox{0.93}{
  \begin{tabular}{lllll}
    \toprule
    Dataset  & 1-mer &+ MoCE &+ Def Conv&+ Noise (MxDNA)  \\
    \midrule
    Average & 88.56 \tiny{\(\pm\) 0.02}  & 88.72  \tiny{\(\pm\) 0.07} &88.86 \tiny{\(\pm\) 0.18}
& 88.89  \tiny{\(\pm\) 0.05}\\
    \midrule
    Mouse Enhancers& 77.56 \tiny{\(\pm\) 0.94}  & 79.66 \tiny{\(\pm\) 0.79}&80.36 \tiny{\(\pm\) 0.39}
&   80.44  \tiny{\(\pm\) 0.63}\\
    Coding vs Intergenomic&95.05 \tiny{\(\pm\) 0.13} &94.64 \tiny{\(\pm\) 0.10}&94.57 \tiny{\(\pm\) 0.43}
&94.78  \tiny{\(\pm\) 0.04}\\
    Human vs Worm&97.52 \tiny{\(\pm\) 0.04}& 97.20 \tiny{\(\pm\) 0.03}&97.31 \tiny{\(\pm\) 0.05}
&97.27  \tiny{\(\pm\) 0.07}\\
    Human Enhancer Cohn&73.70 \tiny{\(\pm\) 0.53}  &74.78 \tiny{\(\pm\) 0.52}& 73.75 \tiny{\(\pm\) 1.10}
& 73.98  \tiny{\(\pm\) 0.40}\\
    Human Enhancer Ensembl&92.79 \tiny{\(\pm\) 0.09} & 92.59 \tiny{\(\pm\) 0.19}&92.78 \tiny{\(\pm\) 0.19}
& 92.73  \tiny{\(\pm\) 0.08}\\
    Human Regulatory&94.03 \tiny{\(\pm\) 0.02}  & 93.93 \tiny{\(\pm\) 0.04}&94.15 \tiny{\(\pm\) 0.11}
& 94.10  \tiny{\(\pm\) 0.12}\\
    Human OCR Ensembl&80.84 \tiny{\(\pm\) 0.50}  &80.27 \tiny{\(\pm\) 0.34}&81.14 \tiny{\(\pm\) 0.08}
&80.62  \tiny{\(\pm\) 0.42}\\
    Human NonTATA Promoters&97.00 \tiny{\(\pm\) 0.05}  & 96.92 \tiny{\(\pm\) 0.32}& 96.81 \tiny{\(\pm\) 0.26}&97.22  \tiny{\(\pm\) 0.23}\\
    \bottomrule
  \end{tabular}}
\end{table}

\begin{table}
  \caption{Nucleotide Transformer Benchmarks. Different components. }
  \label{AblationFullNTCom}
  \centering
  \scalebox{0.93}{
  \begin{tabular}{lllll}
    \toprule
    Dataset  & 1-mer &+ MoCE &+ Def Conv&+ Noise (MxDNA)  \\
    \midrule
    Average &75.07 \tiny{\(\pm\) 0.26} &77.00 \tiny{\(\pm\) 0.05}&77.35 \tiny{\(\pm\) 0.12}& 77.52 \tiny{\(\pm\) 0.18} \\
    \midrule
    Histone Markers Avg. & 63.13 \tiny{\(\pm\) 0.34} & 66.58 \tiny{\(\pm\) 0.11}& 67.02 \tiny{\(\pm\) 0.10}& 67.29 \tiny{\(\pm\) 0.23} \\
    \midrule
    H3& 80.92 \tiny{\(\pm\) 0.85}  & 81.18 \tiny{\(\pm\) 0.52}& 81.65 \tiny{\(\pm\) 0.82}
& 82.14 \tiny{\(\pm\) 0.76}\\
    H3K14ac&62.00 \tiny{\(\pm\) 1.79} &67.68 \tiny{\(\pm\) 0.44}& 66.12 \tiny{\(\pm\) 0.46}
& 68.29 \tiny{\(\pm\)0.65}\\
    H3K36me3&62.59 \tiny{\(\pm\) 1.50} & 66.51 \tiny{\(\pm\) 0.46}&65.26 \tiny{\(\pm\) 1.33}
& 65.46 \tiny{\(\pm\) 1.74}\\
    H3K4me1&51.66\tiny{\(\pm\) 0.57}  &  53.18 \tiny{\(\pm\) 2.31}& 56.14 \tiny{\(\pm\) 1.04}
& 54.97 \tiny{\(\pm\) 1.50}\\
    H3K4me2&49.51 \tiny{\(\pm\) 0.95}&  53.86 \tiny{\(\pm\) 4.01}& 54.13 \tiny{\(\pm\) 1.22}
&  55.30 \tiny{\(\pm\) 0.49}\\
    H3K4me3&54.14 \tiny{\(\pm\) 0.95}  & 63.82 \tiny{\(\pm\) 1.72}& 61.42 \tiny{\(\pm\) 1.58}
& 63.82 \tiny{\(\pm\) 0.92}\\
    H3K79me3&70.36\tiny{\(\pm\) 1.58} & 72.72 \tiny{\(\pm\) 0.71}& 72.88 \tiny{\(\pm\) 0.75}& 73.74 \tiny{\(\pm\) 0.78}\\
    H3K9ac&60.63 \tiny{\(\pm\) 2.73} &63.78 \tiny{\(\pm\) 0.43}& 64.95 \tiny{\(\pm\) 1.57}& 63.15 \tiny{\(\pm\) 0.26}\\
    H4&80.23 \tiny{\(\pm\) 0.79}   &79.96 \tiny{\(\pm\) 0.42}& 81.47 \tiny{\(\pm\) 0.97}& 80.89 \tiny{\(\pm\) 0.23}\\
    H4ac&59.25 \tiny{\(\pm\) 1.36}  &64.09 \tiny{\(\pm\) 0.73}& 66.21 \tiny{\(\pm\) 0.41}& 65.14 \tiny{\(\pm\) 0.23}\\
    \midrule
    Regulatory Annotation Avg. & 85.38 \tiny{\(\pm\) 0.15} & 85.60 \tiny{\(\pm\) 0.32}& 85.71 \tiny{\(\pm\) 0.29}& 85.70 \tiny{\(\pm\) 0.21}\\
    \midrule
    Enhancer&78.93 \tiny{\(\pm\) 0.70} & 79.07 \tiny{\(\pm\) 0.29}& 78.91 \tiny{\(\pm\) 0.29}&  79.73 \tiny{\(\pm\) 0.42}\\
    Enhancer Types&58.90 \tiny{\(\pm\) 0.72} & 60.07 \tiny{\(\pm\) 0.82}& 60.28 \tiny{\(\pm\) 1.20}& 59.79 \tiny{\(\pm\) 0.52}\\
    Promoter All&96.92 \tiny{\(\pm\) 0.08}& 96.92 \tiny{\(\pm\) 0.17}&  96.80 \tiny{\(\pm\) 0.03}& 96.87 \tiny{\(\pm\) 0.10}\\
    Promoter Non-TATA&97.04 \tiny{\(\pm\) 0.05} & 96.71 \tiny{\(\pm\) 0.13}& 96.96 \tiny{\(\pm\) 0.11}& 96.81 \tiny{\(\pm\) 0.13}\\
    Promoter TATA&95.09 \tiny{\(\pm\) 0.34}& 95.25 \tiny{\(\pm\) 0.34}& 95.69 \tiny{\(\pm\) 0.25}& 95.32 \tiny{\(\pm\) 0.17}\\
    \midrule
    Splice Site Annotation Avg. & 97.70 \tiny{\(\pm\) 0.19}&97.39 \tiny{\(\pm\) 0.14}& 97.86 \tiny{\(\pm\) 0.11}& 98.00 \tiny{\(\pm\)0.13}\\
    \midrule
    All&98.07 \tiny{\(\pm\) 0.17} & 98.20 \tiny{\(\pm\) 0.03}&98.13 \tiny{\(\pm\)  0.09}&98.05 \tiny{\(\pm\) 0.13}\\
    Accpetor&97.61 \tiny{\(\pm\) 0.44}  & 97.08 \tiny{\(\pm\) 0.13}& 97.76 \tiny{\(\pm\) 0.21}& 97.68 \tiny{\(\pm\) 0.20}\\
    Donor&97.42 \tiny{\(\pm\) 0.58}&96.90 \tiny{\(\pm\) 0.52} & 97.69 \tiny{\(\pm\) 0.50}& 98.28 \tiny{\(\pm\) 0.07}\\
    \bottomrule
  \end{tabular}}
\end{table}

\subsubsection{Different Tokenization Methods}

Detailed results on each datasets with different tokenization methods are presented in Table~\ref{AblationFullGBTok} and Table~\ref{AblationFullNTTok}. ``1-mer'' stands for single nucleotide. ``Ovlp 6-mer'' stands for overlapping 6-mer. ``Non-Ovlp'' stands for non-overlapping 6-mer. ``BPE'' stands for Byte-pair Encoding with a vocabulary size of 4096 borrowed from DNABERT2. All models are trained for 100k steps with same backbone but different tokenization methods.

\subsubsection{Different Components}

Detailed results on each datasets with components added from the single nucleotide tokenization baseline are presented in Table~\ref{AblationFullGBCom} and Table~\ref{AblationFullNTCom}. ``1-mer'' stands for the baseline. ``+ MoCE'' stands for adding the sparse Mixture of Convolution Experts.``+ Def Conv'' stands for adding the deformable convolution block. ``+ Noise'' stands for adding the multiplicative jitter noise. These components are added successively, finally equivalent to MxDNA. All models are trained for 100k steps with same backbone and components added sequentially.

\subsection{Visualization Details}

The BPE tokenizer used here is directly borrowed from DNABERT2 with a vocabulary of size 4096. The visualization methods for traditional tokenization methods is straightforward. Below are the visualization details for MxDNA.

\label{AppendixVis}
\paragraph{Sample Level:} For MxDNA, we perform a forward process of the model using the sequence as input, extracting the Mask of basic units existence \(M\), Offsets \(\Delta P\) and Modulation factors \(\Delta M\) of the deformable convolution. First, we colour the recognized basic units based on \(M\). Then, we determine the distal relations using \(\Delta P\) and \(\Delta M\). Specifically, a distal relation is considered to be visualized only if the the product of the corresponding modulation weight and the bilinearly interpolated offset weight exceeds one. Eventually, a final learnt token is made up of a group of related basic units and coloured by the colour of the central basic unit. 

\paragraph{Dataset Level:} For dataset level, we first finetune the MxDNA model on downstream datasets and use the refined models to generate the Mask of basic units existence \(M\) for all samples in the dataset. We then calculate the proportion of the lengths of each recognized basic units as indicated by \(M\). Specifically, we finetune MxDNA on H3, Enhancer, Promoter All and Splice Site All in Nucleotide Transformer Benchmarks for Histone Marker, Enhancer, Promoter and Splice Site respectively.

\paragraph{Token Embedding Analysis:} For the token embedding analysis, we utilize various pretrained models to embed sequences with different functions and analyse the output embeddings at the token level. Initially, we perform principal component analysis to reduce the dimensionality to one hundred, which facilitates the visualization process. Subsequently, we employ t-SNE to visualize these token embeddings in a two-dimensional space. Specifically, we use the H3, Enhancer, Promoter All, and Splice Site All sequences from the test set of the Nucleotide Transformer Benchmarks to represent Histone Marker, Enhancer, Promoter, and Splice Site, respectively.

\subsection{Computational Resources}

\label{AppendixCompute}

We train and evaluate the models on NVIDIA RTX 3090 and NVIDIA A100 GPUs. The pretraining of MxDNA takes around 3 days for 500k steps using 4 A100 GPUs. Finetuing MxDNA on all the downstream tasks takes approximately 1.5 Days using 1 A100 GPU. This is true for other BERT-like foundation models with around 100M parameters.

The detailed computational costs of the models (averaged across 5 samples of sequence length of 510) are outlined in Table~\ref{tab:computation}. The integration of a mixture of convolution experts and the deformable convolution introduces an increased computational overhead initially due to the \(O(l \log(l))\) time complexity of the learned tokenization mechanism (where \(l\) represents the number of nucleotides). This complexity is mitigated by the substantial reduction in sequence length after tokenization, which decreases the number of tokens processed by subsequent transformer layers.

\begin{table}
\centering
\caption{Comparison of various models based on their computational complexity.}
\label{tab:computation}
\begin{tabular}{lrrrr}
\toprule
\textbf{Model} & Flops (G) & Macs (G) & Parameters (M) & Number of Tokens \\
\midrule
DNABERT2                    & 24.80 & 12.39 & 117.07 & 104.2 \\
NTv2 100M                   & 16.63 & 8.31  & 97.89  & 86 \\
DNABERT                     & 99.48 & 49.70 & 89.20  & 507 \\
HyenaDNA tiny d256          & 1.67  & 0.832 & 1.64   & 511 \\
HyenaDNA tiny               & 0.441 & 0.219 & 0.436  & 511 \\
MxDNA                       & 35.94 & 17.93 & 100.09 & 512 $\rightarrow$ 101.6 \\
Learnt Tokenization Module  & 0.914 & 0.446 & 11.69  & 512 $\rightarrow$ 101.6 \\
Single Nucleotide Baseline  & 94.85 & 47.38 & 92.95  & 512 \\
\bottomrule
\end{tabular}
\end{table}

\subsection{Assets}
 The assets used in this work along with their licenses including data, pretrained weights, benchmarks, libraries, and software are presented in Table~\ref{LicenseTable}.
 
\label{AppendixLicense}

\begin{table}
  \caption{Assets used in this work}
  \label{LicenseTable}
  \centering
  \begin{tabular}{ll}
    \toprule
    Asset & License  \\
    \midrule
    GRCh38~\cite{HG38} & CC BY 4.0 \\
    Genomic Benchmarks~\cite{grevsova2023genomic} & Apache-2.0  \\
    Nucleotide Transformer~\cite{dalla2023nucleotide} & CC BY-NC-SA 4.0 \\
    DNABERT~\cite{ji2021dnabert} & Apache-2.0 \\
    DNABERT2~\cite{zhou2023dnabert} & Apache-2.0 \\
    HyenaDNA~\cite{nguyen2024hyenadna}& Apache-2.0 \\
    FlashAttention~\cite{dao2022flashattention,dao2023flashattention} & BSD-3-Clause  \\
    Pytorch~\cite{Ansel_PyTorch_2_Faster_2024}& BSD-3-Clause  \\
    Pytorch Lightning~\cite{Falcon_PyTorch_Lightning_2019} & Apache-2.0 \\
    Huggingface~\cite{wolf-etal-2020-transformers} & Apache-2.0 \\
    Pybind11~\cite{pybind11} & BSD-3-Clause \\
    Scikit-Learn~\cite{scikit-learn} & BSD-3-Clause  \\
    Numpy~\cite{harris2020array} & BSD-3-Clause  \\
    Matplotlib~\cite{Hunter:2007} & \href{https://matplotlib.org/stable/project/license.html}{Matplotlib License}\\
    Seaborn~\cite{Waskom2021} & Apache-2.0  \\
    \bottomrule
  \end{tabular}
\end{table}

\end{document}